\documentclass[longauth]{aa}
\usepackage{graphicx}
\usepackage{txfonts}
\usepackage{hyperref}
\newcommand {\bc}{\begin {center}}
\newcommand {\ec}{\end {center}}
\newcommand {\be}{\begin {equation}}
\newcommand {\ee}{\end {equation}}
\newcommand {\beq}{\begin {eqnarray}}
\newcommand {\eeq}{\end {eqnarray}}

\newcommand {\lum}{erg~s$^{-1}$}

\newcommand {\ixpe}{\textit{IXPE}\xspace}

\newcommand {\gro}{\mbox{GRO~J1008$-$57}\xspace}

\begin{document}

\title{X-ray pulsar GRO~J1008$-$57 as an orthogonal rotator}

\author{\small 
Sergey~S.~Tsygankov \inst{\ref{in:UTU}}
 \and Victor~Doroshenko \inst{\ref{in:Tub}}
\and Alexander~A.~Mushtukov \inst{\ref{in:Oxford}}
\and Juri~Poutanen \inst{\ref{in:UTU}}
\and Alessandro~Di~Marco \inst{\ref{in:INAF-IAPS}}
\and Jeremy~Heyl \inst{\ref{in:UBC}}
\and Fabio~La~Monaca \inst{\ref{in:INAF-IAPS}} 
\and Sofia~V.~Forsblom \inst{\ref{in:UTU}}
\and Christian~Malacaria \inst{\ref{in:ISSI}}
\and Herman~L.~Marshall \inst{\ref{in:MIT}}
\and Valery~F.~Suleimanov \inst{\ref{in:Tub}}
\and Jiri~Svoboda \inst{\ref{in:CAS-ASU}}
\and Roberto~Taverna \inst{\ref{in:UniPD}}  
\and Francesco~Ursini \inst{\ref{in:UniRoma3}} 
\and Iv\'an~Agudo \inst{\ref{in:CSIC-IAA}}
\and Lucio~A.~Antonelli \inst{\ref{in:INAF-OAR},\ref{in:ASI-SSDC}} 
\and Matteo~Bachetti \inst{\ref{in:INAF-OAC}} 
\and Luca~Baldini  \inst{\ref{in:INFN-PI},      \ref{in:UniPI}} 
\and Wayne~H.~Baumgartner  \inst{\ref{in:NASA-MSFC}} 
\and Ronaldo~Bellazzini  \inst{\ref{in:INFN-PI}} 
\and Stefano~Bianchi \inst{\ref{in:UniRoma3}}  
\and Stephen~D.~Bongiorno \inst{\ref{in:NASA-MSFC}} 
\and Raffaella~Bonino  \inst{\ref{in:INFN-TO},\ref{in:UniTO}}
\and Alessandro~Brez  \inst{\ref{in:INFN-PI}} 
\and Niccol\`{o}~Bucciantini 
\inst{\ref{in:INAF-Arcetri},\ref{in:UniFI},\ref{in:INFN-FI}} 
\and Fiamma~Capitanio \inst{\ref{in:INAF-IAPS}}
\and Simone~Castellano \inst{\ref{in:INFN-PI}}  
\and Elisabetta~Cavazzuti \inst{\ref{in:ASI}} 
\and Chien-Ting~Chen \inst{\ref{in:USRA-MSFC}}
\and Stefano~Ciprini \inst{\ref{in:INFN-Roma2},\ref{in:ASI-SSDC}}
\and Enrico~Costa \inst{\ref{in:INAF-IAPS}} 
\and Alessandra~De~Rosa \inst{\ref{in:INAF-IAPS}} 
\and Ettore~Del~Monte \inst{\ref{in:INAF-IAPS}} 
\and Laura~Di~Gesu \inst{\ref{in:ASI}} 
\and Niccol\`{o}~Di~Lalla \inst{\ref{in:Stanford}}
\and Immacolata~Donnarumma \inst{\ref{in:ASI}}
\and Michal~Dov\v{c}iak \inst{\ref{in:CAS-ASU}}
\and Steven~R.~Ehlert \inst{\ref{in:NASA-MSFC}}  
\and Teruaki~Enoto \inst{\ref{in:RIKEN}}
\and Yuri~Evangelista \inst{\ref{in:INAF-IAPS}}
\and Sergio~Fabiani \inst{\ref{in:INAF-IAPS}}
\and Riccardo~Ferrazzoli \inst{\ref{in:INAF-IAPS}} 
\and Javier~A.~Garcia \inst{\ref{in:Caltech}}
\and Shuichi~Gunji\inst{\ref{in:Yamagata}} 
\and Kiyoshi~Hayashida \inst{\ref{in:Osaka}}\thanks{Deceased.} 
\and Wataru~Iwakiri \inst{\ref{in:Chiba}} 
\and Svetlana~G.~Jorstad \inst{\ref{in:BU},\ref{in:SPBU}} 
\and Philip~Kaaret \inst{\ref{in:NASA-MSFC}}  
\and Vladimir~Karas \inst{\ref{in:CAS-ASU}}
\and Fabian~Kislat \inst{\ref{in:UNH}} 
\and Takao~Kitaguchi  \inst{\ref{in:RIKEN}} 
\and Jeffery~J.~Kolodziejczak \inst{\ref{in:NASA-MSFC}} 
\and Henric~Krawczynski  \inst{\ref{in:WUStL}}
\and Luca~Latronico  \inst{\ref{in:INFN-TO}} 
\and Ioannis~Liodakis \inst{\ref{in:FINCA}}
\and Simone~Maldera \inst{\ref{in:INFN-TO}}  
\and Alberto~Manfreda \inst{\ref{INFN-NA}}
\and Fr\'{e}d\'{e}ric~Marin \inst{\ref{in:Strasbourg}} 
\and Andrea~Marinucci \inst{\ref{in:ASI}} 
\and Alan~P.~Marscher \inst{\ref{in:BU}} 
\and Francesco~Massaro \inst{\ref{in:INFN-TO},\ref{in:UniTO}} 
\and Giorgio~Matt  \inst{\ref{in:UniRoma3}}  
\and Ikuyuki~Mitsuishi \inst{\ref{in:Nagoya}} 
\and Tsunefumi~Mizuno \inst{\ref{in:Hiroshima}} 
\and Fabio~Muleri \inst{\ref{in:INAF-IAPS}} 
\and Michela~Negro \inst{\ref{in:UMBC},\ref{in:NASA-GSFC},\ref{in:CRESST}} 
\and Chi-Yung~Ng \inst{\ref{in:HKU}}
\and Stephen~L.~O'Dell \inst{\ref{in:NASA-MSFC}}  
\and Nicola~Omodei \inst{\ref{in:Stanford}}
\and Chiara~Oppedisano \inst{\ref{in:INFN-TO}}  
\and Alessandro~Papitto \inst{\ref{in:INAF-OAR}}
\and George~G.~Pavlov \inst{\ref{in:PSU}}
\and Abel~L.~Peirson \inst{\ref{in:Stanford}}
\and Matteo~Perri \inst{\ref{in:ASI-SSDC},\ref{in:INAF-OAR}}
\and Melissa~Pesce-Rollins \inst{\ref{in:INFN-PI}} 
\and Pierre-Olivier~Petrucci \inst{\ref{in:Grenoble}} 
\and Maura~Pilia \inst{\ref{in:INAF-OAC}} 
\and Andrea~Possenti \inst{\ref{in:INAF-OAC}} 
\and Simonetta~Puccetti \inst{\ref{in:ASI-SSDC}}
\and Brian~D.~Ramsey \inst{\ref{in:NASA-MSFC}}  
\and John~Rankin \inst{\ref{in:INAF-IAPS}} 
\and Ajay~Ratheesh \inst{\ref{in:INAF-IAPS}} 
\and Oliver~J.~Roberts \inst{\ref{in:USRA-MSFC}}
\and Roger~W.~Romani \inst{\ref{in:Stanford}}
\and Carmelo~Sgr\`{o} \inst{\ref{in:INFN-PI}}  
\and Patrick~Slane \inst{\ref{in:CfA}}  
\and Paolo~Soffitta \inst{\ref{in:INAF-IAPS}} 
\and Gloria~Spandre \inst{\ref{in:INFN-PI}} 
\and Douglas~A.~Swartz \inst{\ref{in:USRA-MSFC}}
\and Toru~Tamagawa \inst{\ref{in:RIKEN}}
\and Fabrizio~Tavecchio \inst{\ref{in:INAF-OAB}}
\and Yuzuru~Tawara \inst{\ref{in:Nagoya}}
\and Allyn~F.~Tennant \inst{\ref{in:NASA-MSFC}}  
\and Nicholas~E.~Thomas \inst{\ref{in:NASA-MSFC}}  
\and Francesco~Tombesi  \inst{\ref{in:UniRoma2},\ref{in:INFN-Roma2},\ref{in:UMd}}
\and Alessio~Trois \inst{\ref{in:INAF-OAC}}
\and Roberto~Turolla \inst{\ref{in:UniPD},\ref{in:MSSL}}
\and Jacco~Vink \inst{\ref{in:Amsterdam}}
\and Martin~C.~Weisskopf \inst{\ref{in:NASA-MSFC}} 
\and Kinwah~Wu \inst{\ref{in:MSSL}}
\and Fei~Xie \inst{\ref{in:GSU},\ref{in:INAF-IAPS}}
\and Silvia Zane  \inst{\ref{in:MSSL}}
          }
          
\institute{Department of Physics and Astronomy, FI-20014 University of Turku,  Finland \label{in:UTU} \\ \email{sergey.tsygankov@utu.fi}
\and
Institut f\"ur Astronomie und Astrophysik, Universit\"at T\"ubingen, Sand 1, D-72076 T\"ubingen, Germany \label{in:Tub}
\and 
Astrophysics, Department of Physics, University of Oxford, Denys Wilkinson Building, Keble Road, Oxford OX1 3RH, UK \label{in:Oxford}
\and 
INAF Istituto di Astrofisica e Planetologia Spaziali, Via del Fosso del Cavaliere 100, 00133 Roma, Italy \label{in:INAF-IAPS}
\and 
University of British Columbia, Vancouver, BC V6T 1Z4, Canada \label{in:UBC}
\and 
International Space Science Institute, Hallerstrasse 6, 3012 Bern, Switzerland \label{in:ISSI}
\and 
MIT Kavli Institute for Astrophysics and Space Research, Massachusetts Institute of Technology, 77 Massachusetts Avenue, Cambridge, MA 02139, USA \label{in:MIT}
\and 
Astronomical Institute of the Czech Academy of Sciences, Boční II 1401/1, 14100 Praha 4, Czech Republic \label{in:CAS-ASU}
\and 
Dipartimento di Fisica e Astronomia, Universit\`{a} degli Studi di Padova, Via Marzolo 8, 35131 Padova, Italy \label{in:UniPD}
\and 
Dipartimento di Matematica e Fisica, Universit\`a degli Studi Roma Tre, via della Vasca Navale 84, 00146 Roma, Italy  \label{in:UniRoma3}
\and 
Instituto de Astrof\'{i}sicade Andaluc\'{i}a -- CSIC, Glorieta de la Astronom\'{i}a s/n, 18008 Granada, Spain \label{in:CSIC-IAA}
\and 
INAF Osservatorio Astronomico di Roma, Via Frascati 33, 00040 Monte Porzio Catone (RM), Italy \label{in:INAF-OAR}  
\and 
Space Science Data Center, Agenzia Spaziale Italiana, Via del Politecnico snc, 00133 Roma, Italy \label{in:ASI-SSDC}
 \and
INAF Osservatorio Astronomico di Cagliari, Via della Scienza 5, 09047 Selargius (CA), Italy  \label{in:INAF-OAC}
\and 
Istituto Nazionale di Fisica Nucleare, Sezione di Pisa, Largo B. Pontecorvo 3, 56127 Pisa, Italy \label{in:INFN-PI}
\and  
Dipartimento di Fisica, Universit\`{a} di Pisa, Largo B. Pontecorvo 3, 56127 Pisa, Italy \label{in:UniPI} 
\and 
NASA Marshall Space Flight Center, Huntsville, AL 35812, USA \label{in:NASA-MSFC}
\and  
Istituto Nazionale di Fisica Nucleare, Sezione di Torino, Via Pietro Giuria 1, 10125 Torino, Italy  \label{in:INFN-TO}      
\and  
Dipartimento di Fisica, Universit\`{a} degli Studi di Torino, Via Pietro Giuria 1, 10125 Torino, Italy \label{in:UniTO} 
\and   
INAF Osservatorio Astrofisico di Arcetri, Largo Enrico Fermi 5, 50125 Firenze, Italy 
\label{in:INAF-Arcetri} 
\and  
Dipartimento di Fisica e Astronomia, Universit\`{a} degli Studi di Firenze, Via Sansone 1, 50019 Sesto Fiorentino (FI), Italy \label{in:UniFI} 
\and   
Istituto Nazionale di Fisica Nucleare, Sezione di Firenze, Via Sansone 1, 50019 Sesto Fiorentino (FI), Italy \label{in:INFN-FI}
\and 
Agenzia Spaziale Italiana, Via del Politecnico snc, 00133 Roma, Italy \label{in:ASI}
\and 
Science and Technology Institute, Universities Space Research Association, Huntsville, AL 35805, USA \label{in:USRA-MSFC}
\and 
Istituto Nazionale di Fisica Nucleare, Sezione di Roma ``Tor Vergata'', Via della Ricerca Scientifica 1, 00133 Roma, Italy 
 \label{in:INFN-Roma2}
\and 
Department of Physics and Kavli Institute for Particle Astrophysics and Cosmology, Stanford University, Stanford, California 94305, USA  \label{in:Stanford}
\and 
RIKEN Cluster for Pioneering Research, 2-1 Hirosawa, Wako, Saitama 351-0198, Japan \label{in:RIKEN}
\and 
California Institute of Technology, Pasadena, CA 91125, USA \label{in:Caltech}
\and 
Yamagata University,1-4-12 Kojirakawa-machi, Yamagata-shi 990-8560, Japan \label{in:Yamagata}
\and 
Osaka University, 1-1 Yamadaoka, Suita, Osaka 565-0871, Japan \label{in:Osaka}
\and 
International Center for Hadron Astrophysics, Chiba University, Chiba 263-8522, Japan \label{in:Chiba}
\and
Institute for Astrophysical Research, Boston University, 725 Commonwealth Avenue, Boston, MA 02215, USA \label{in:BU} 
\and 
Department of Astrophysics, St. Petersburg State University, Universitetsky pr. 28, Petrodvoretz, 198504 St. Petersburg, Russia \label{in:SPBU} 
\and 
Department of Physics and Astronomy and Space Science Center, University of New Hampshire, Durham, NH 03824, USA \label{in:UNH} 
\and 
Physics Department and McDonnell Center for the Space Sciences, Washington University in St. Louis, St. Louis, MO 63130, USA \label{in:WUStL}
\and 
Finnish Centre for Astronomy with ESO,  20014 University of Turku, Finland \label{in:FINCA}
\and 
Istituto Nazionale di Fisica Nucleare, Sezione di Napoli, Strada Comunale Cinthia, 80126 Napoli, Italy \label{INFN-NA}
\and 
Universit\'{e} de Strasbourg, CNRS, Observatoire Astronomique de Strasbourg, UMR 7550, 67000 Strasbourg, France \label{in:Strasbourg} 
\newpage 
\and 
Graduate School of Science, Division of Particle and Astrophysical Science, Nagoya University, Furo-cho, Chikusa-ku, Nagoya, Aichi 464-8602, Japan \label{in:Nagoya}
\and 
Hiroshima Astrophysical Science Center, Hiroshima University, 1-3-1 Kagamiyama, Higashi-Hiroshima, Hiroshima 739-8526, Japan \label{in:Hiroshima}
\and  University of Maryland, Baltimore County, Baltimore, MD 21250, USA \label{in:UMBC}
\and NASA Goddard Space Flight Center, Greenbelt, MD 20771, USA  \label{in:NASA-GSFC}
\and Center for Research and Exploration in Space Science and Technology, NASA/GSFC, Greenbelt, MD 20771, USA  \label{in:CRESST}
\and 
Department of Physics, University of Hong Kong, Pokfulam, Hong Kong \label{in:HKU}
\and 
Department of Astronomy and Astrophysics, Pennsylvania State University, University Park, PA 16801, USA \label{in:PSU}
\and 
Universit\'{e} Grenoble Alpes, CNRS, IPAG, 38000 Grenoble, France \label{in:Grenoble}
\and 
Center for Astrophysics, Harvard \& Smithsonian, 60 Garden St, Cambridge, MA 02138, USA \label{in:CfA} 
\and 
INAF Osservatorio Astronomico di Brera, via E. Bianchi 46, 23807 Merate (LC), Italy \label{in:INAF-OAB}
\and
Dipartimento di Fisica, Universit\`{a} degli Studi di Roma ``Tor Vergata'', Via della Ricerca Scientifica 1, 00133 Roma, Italy \label{in:UniRoma2}
\and
Department of Astronomy, University of Maryland, College Park, Maryland 20742, USA \label{in:UMd}
\and 
Mullard Space Science Laboratory, University College London, Holmbury St Mary, Dorking, Surrey RH5 6NT, UK \label{in:MSSL}
\and 
Anton Pannekoek Institute for Astronomy \& GRAPPA, University of Amsterdam, Science Park 904, 1098 XH Amsterdam, The Netherlands  \label{in:Amsterdam}
\and 
Guangxi Key Laboratory for Relativistic Astrophysics, School of Physical Science and Technology, Guangxi University, Nanning 530004, China \label{in:GSU}
}
          
\titlerunning{X-ray polarimetry of X-ray pulsar GRO~J1008--57}
\authorrunning{S.~Tsygankov et al.}

\date{2023}

\abstract{
X-ray polarimetry is a unique way to probe 
the geometrical configuration of highly magnetized accreting neutron stars (X-ray pulsars). GRO~J1008$-$57 is the first transient X-ray pulsar observed at two different flux levels by the Imaging X-ray Polarimetry Explorer (IXPE) during its outburst in November 2022. We find the polarization properties of GRO~J1008$-$57 to be independent of its luminosity, with the polarization degree varying between nondetection and about 15\% over the pulse phase. Fitting the phase-resolved spectro-polarimetric data with the rotating vector model allowed us to estimate the pulsar inclination (130\degr, which is in good agreement with the orbital inclination), the position angle (75\degr) of the pulsar spin axis, and the magnetic obliquity ($\sim74$\degr). This makes GRO~J1008$-$57 the first confidently identified nearly orthogonal rotator among X-ray pulsars. We discuss our results in the context of the neutron star atmosphere models and theories of the axis alignment of accreting pulsars. 
}


\maketitle
%
\section{Introduction} 
\label{sec:intro}

The physics of the interaction of astrophysical plasmas with ultrastrong magnetic and radiation fields in the vicinity of neutron stars (NSs) is reflected in the observed properties of accreting X-ray pulsars \citep[XRPs; see][for a recent review]{Mushtukov22}. Therefore, the analysis of observational data can, in principle, be used to study the complex interplay of several physical processes defining such interactions. However, the sheer complexity of the problem together with the large uncertainty in the basic geometry of pulsar emission regions have so far limited the potential of X-ray pulsars as laboratories for studying physics under extreme conditions. The situation might change with the launch of the first imaging space X-ray polarimeter, the \textit{Imaging X-ray Polarimeter Explorer} \citep[\ixpe,][]{Weisskopf2022}, which has opened a new observational window onto X-ray pulsars and is considered to be a unique tool with which to break several model degeneracies through independent constraints on the geometrical parameters of the system. 
During the first year of operation in orbit, \ixpe observed several XRPs, yielding results that are at surprising variance with pre-launch theoretical predictions \citep{Doroshenko22,2022ApJ...941L..14T}. In particular, the emission of XRPs was expected to be strongly polarized (up to 80\%; see, e.g., \citealt{1988ApJ...324.1056M,2021MNRAS.501..109C,2022ApJ...939...67B}), and a significantly lower polarization degree (PD) was expected to only be theoretically possible at low accretion rates due to the inverse temperature profile in the atmosphere of an accreting NS \citep{2019MNRAS.483..599G,2021MNRAS.503.5193M}. However, it was found that even bright XRPs (with luminosities exceeding $10^{37}$~\lum) show PDs of well below 20\%, even in the phase-resolved data \citep{Doroshenko22,2022ApJ...940...70M,2022ApJ...941L..14T}. 

The cause at the basis of this discrepancy is unclear and can be related not only to the physics within the emission region, but also to the potential complexity of the geometry of the emission region itself, and of the accretion flow, which may affect the observed polarization signal. Different sites of the polarized emission in the NS vicinity were discussed by \cite{2022ApJ...941L..14T}, who proposed, in particular, (i) intrinsic polarization from polar hotspots, (ii) reflection of the emission by the NS surface, (iii) reflection by the accretion curtain, (iv) reflection by the accretion disk, (v) scattering by the stellar wind, and (vi) reflection by the optical companion.  Therefore, to exclude the geometrical factor, probing different scenarios of polarization production in the NS atmosphere requires multiple observations of the same source at different mass-accretion rates.

\begin{figure*}
\centering
\includegraphics[width=0.97\linewidth]{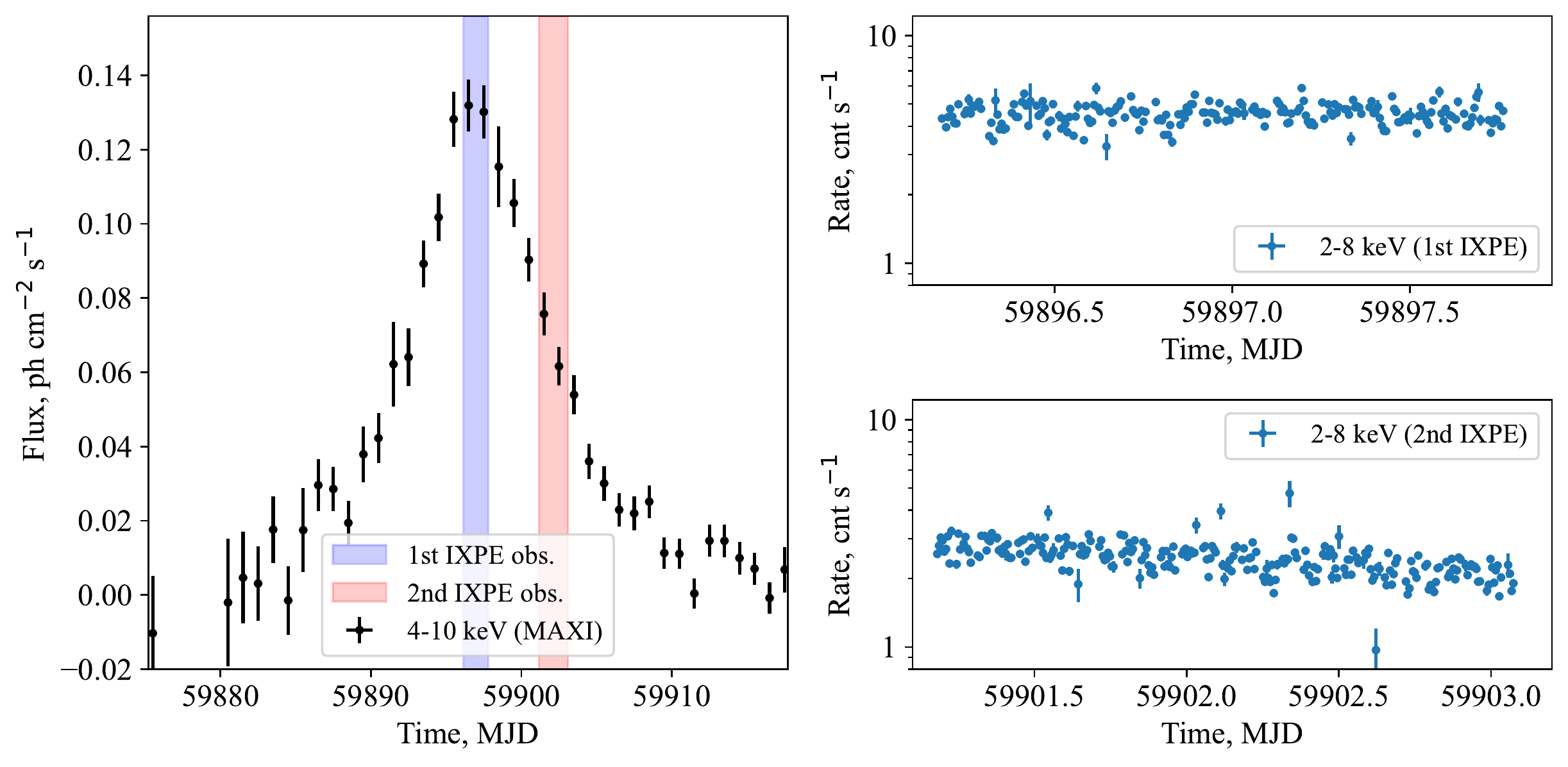}
\caption{Light curve of \gro during the November 2022 outburst. {\it Left:} Long-term light curve in the 4--10 keV energy band obtained by the MAXI all-sky monitor (black points). Blue and red shaded stripes show the time periods of the first and second \ixpe observations, respectively. 
{\it Right:} Light curves of \gro in the 2--8 keV energy band summed over three modules of \ixpe during the first (top) and second (bottom) observation. 
 }
 \label{fig:ixpe-lc}
\end{figure*}

The most obvious candidates for such an observational campaign are transient XRPs with a Be optical companion \citep[Be/XRP; see][for a review]{2011Ap&SS.332....1R} that exhibit regular outbursts every orbital cycle due to enhanced accretion rate at around the time of periastron passage. One of the most predictable and best-studied Be/XRPs is \gro \citep{2013A&A...555A..95K}. This object was discovered in 1993 by the BATSE instrument on board the {\it Compton Gamma-Ray Observatory} as a transient XRP with a spin period of $93.587\pm0.005$\,s \citep{1993IAUC.5836....1S}. \gro\ shows giant Type II and Type I (associated with the periastron passage) outbursts, which are related to the Be type of the optical companion (B1-B2\,Ve star; \citealt{2007MNRAS.378.1427C}).  The typical peak luminosity of the source during the Type I outburst is about 10$^{37}$~\lum, with Type II outbursts being several times brighter, reaching $\sim5\times10^{37}$~\lum. The distance to \gro\ was recently updated using the \textit{Gaia} data to $3.6\pm0.2$~kpc \citep{2022A&A...665A..31F}. The orbital parameters of the binary system are known as well: the orbital period $P_{\rm orb}=249.48\pm0.04$~d, the projected semi-major axis $a_{\rm x}$~sin~$i=530\pm60$~lt~s, the longitude of periastron $\omega=-26\pm8$~deg, and the eccentricity $e = 0.68\pm0.02$ \citep{2007MNRAS.378.1427C, 2013A&A...555A..95K}. 

The broadband energy spectrum of \gro is known to depend on the source luminosity and exhibits a double-hump structure at the lowest fluxes \citep{2021ApJ...912...17L}, which is typical of low-luminosity XRPs \citep[see, e.g.,][]{2019MNRAS.483L.144T,2019MNRAS.487L..30T}. The cyclotron absorption feature at $\sim 88$ keV in the energy spectrum of the source was first discovered in the {\it CGRO}/OSSE data \citep{1999ApJ...512..920S} and was later confirmed by \textit{Suzaku} and \textit{NuSTAR} as the fundamental at $E_{\rm cyc} = 75-78$ keV \citep{2013ATel.4759....1Y,2014ApJ...792..108B} and by {\it Insight-HXMT} at $E_{\rm cyc} = 90.3$ keV \citep{2020ApJ...899L..19G}. This makes \gro an XRP with one of the strongest confirmed  magnetic fields, of around $10^{13}$~G. 
Recently, it was shown that even between periastron passages, the NS continues to steadily accrete matter from the recombined accretion disc, which emits at a level of $\sim10^{34}-10^{35}$~\lum\ \citep{2017A&A...608A..17T,2017MNRAS.470..126T}.

Here, we present the results of the analysis of \gro observations performed by \ixpe during the source periastron passage in two different luminosity states.  First, we describe the observations and the data-reduction procedures in Sect.~\ref{sec:data}. The results are  presented in Sect.~\ref{sec:res}.  We then discuss possible sources of the observed polarization and the geometry of the pulsar in Sect.~\ref{sec:discussion}, and finally summarize our results and present conclusions in Sect.~\ref{sec:sum}. 


\section{Data} 
\label{sec:data}

The \ixpe is a NASA mission in partnership with the Italian Space Agency, and was launched on 2021 December 9. It consists of three identical grazing incidence telescopes, providing imaging polarimetry over the 2--8 keV energy band with a time resolution of the order of 10~$\mu$s. Each telescope comprises an X-ray mirror assembly and a polarization-sensitive detector unit (DU) equipped with a gas-pixel detector \citep{2021AJ....162..208S,2021APh...13302628B}. A detailed description of the observatory and its performance is given in \citet{Weisskopf2022}.

\ixpe observed \gro\ twice during the same Type I outburst over the periods of 2022 Nov 13--14 and Nov 18--20 with a total effective exposure of $\simeq$85~ks and $\simeq$102~ks, respectively. 
Level 2 data were processed with the {\sc ixpeobssim} package \citep{Baldini2022} version 30.2.1\footnote{\url{https://github.com/lucabaldini/ixpeobssim}} using the Calibration database released on 2022 November 17. 

Source photons were collected from a circular region with radius $R_{\rm src}$ of 60\arcsec\ centered on the source position. Following the prescription by \citet{Di_Marco_2023} for the  sources with high count rate ($\gtrsim2$~cnt s$^{-1}$), the background has not been subtracted from the data, because it is negligible.
For the timing analysis, the photon arrival times were corrected to the Solar System barycenter using the standard {\tt barycorr} tool from the {\sc ftools} package, with the effects of binary motion also taken into account using the orbital parameters from \citet{2013A&A...555A..95K}. 

For the spectro-polarimetric analysis, the flux (Stokes parameter $I$) energy spectra have been binned to have at least 30 counts per energy channel.  The same energy binning has also been applied to the spectra of the Stokes parameters $Q$ and $U$. Taking into account the high number of source counts and low background level, the unweighted approach has been applied.
All the spectra were fitted using the {\sc xspec} package version 12.12.1 \citep{Arn96} and version 12 of the instrument response functions, and applying the $\chi^2$ statistic. 
The uncertainties are given at the 68.3\% confidence level unless stated otherwise.
 
\begin{figure*}
\centering
\includegraphics[width=0.95\linewidth]{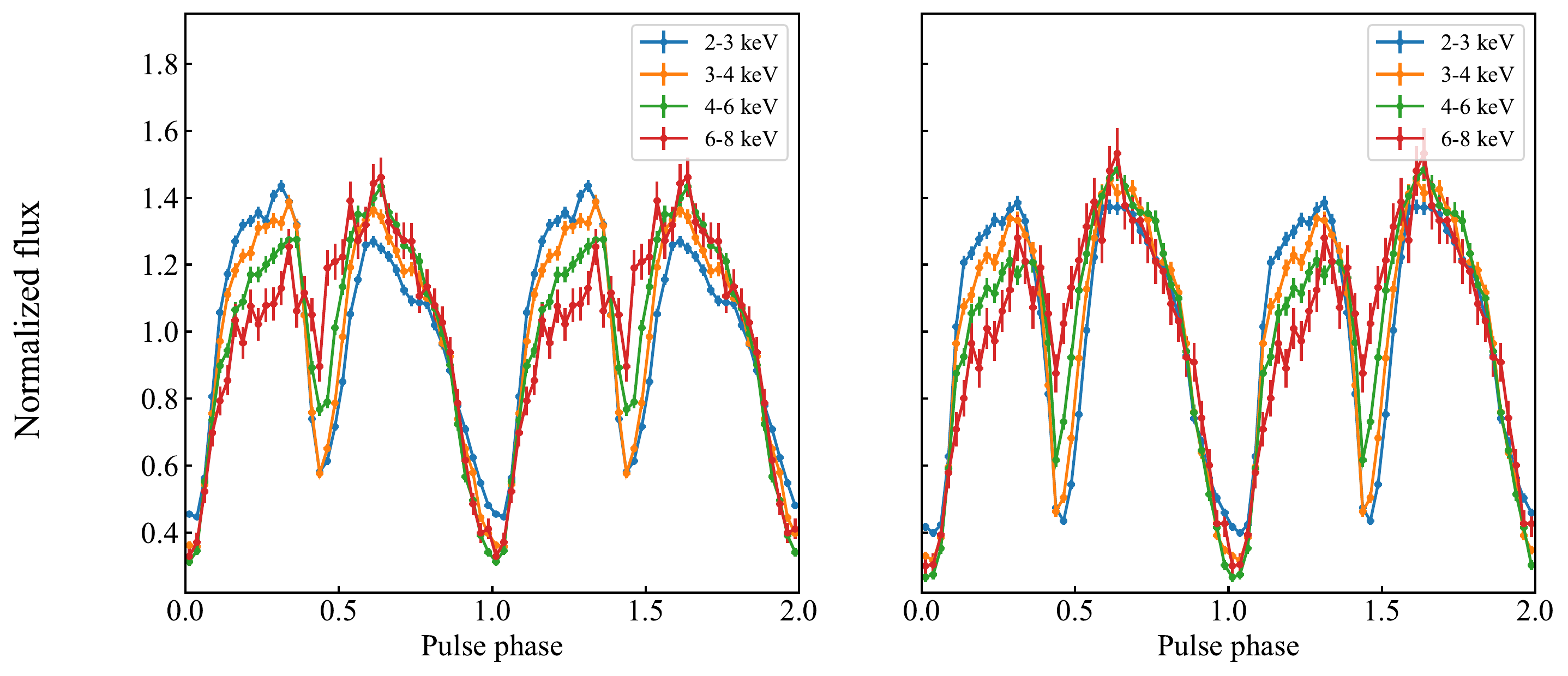}
\caption{Pulse profile of \gro in different energy bands as seen by \ixpe in the bright (left) and low (right) states. Data from the three telescopes were combined. 
The zero phase was chosen to coincide for both observations using cross-correlation of the profiles. 
}
 \label{fig:ixpe-pprof}
\end{figure*}

\section{Results} 
\label{sec:res}

\subsection{Timing analysis}
\label{sec:time}
The long-term light curve of \gro obtained with the MAXI all-sky monitor  \citep{Matsuoka09}\footnote{\url{http://maxi.riken.jp/star_data/J1009-582/J1009-582.html}} is shown in Figure~\ref{fig:ixpe-lc} along with source light curves as observed by \ixpe during the two observations,
which differ by a factor of $\sim$2 in flux.  In the low state, the observed luminosity in the 2--8 keV range was about $L_{\rm low}=8.6\times10^{35}$~erg~s$^{-1}$, while the bright state had $L_{\rm high}=1.6\times10^{36}$~erg~s$^{-1}$. 
The source did not exhibit any significant variability within either individual observation, which allowed us to average all the available data. To study the effect of different mass-accretion rates on the polarization properties, the two observations were first analyzed independently. 

\begin{figure*}
\centering
\includegraphics[width=0.45\linewidth]{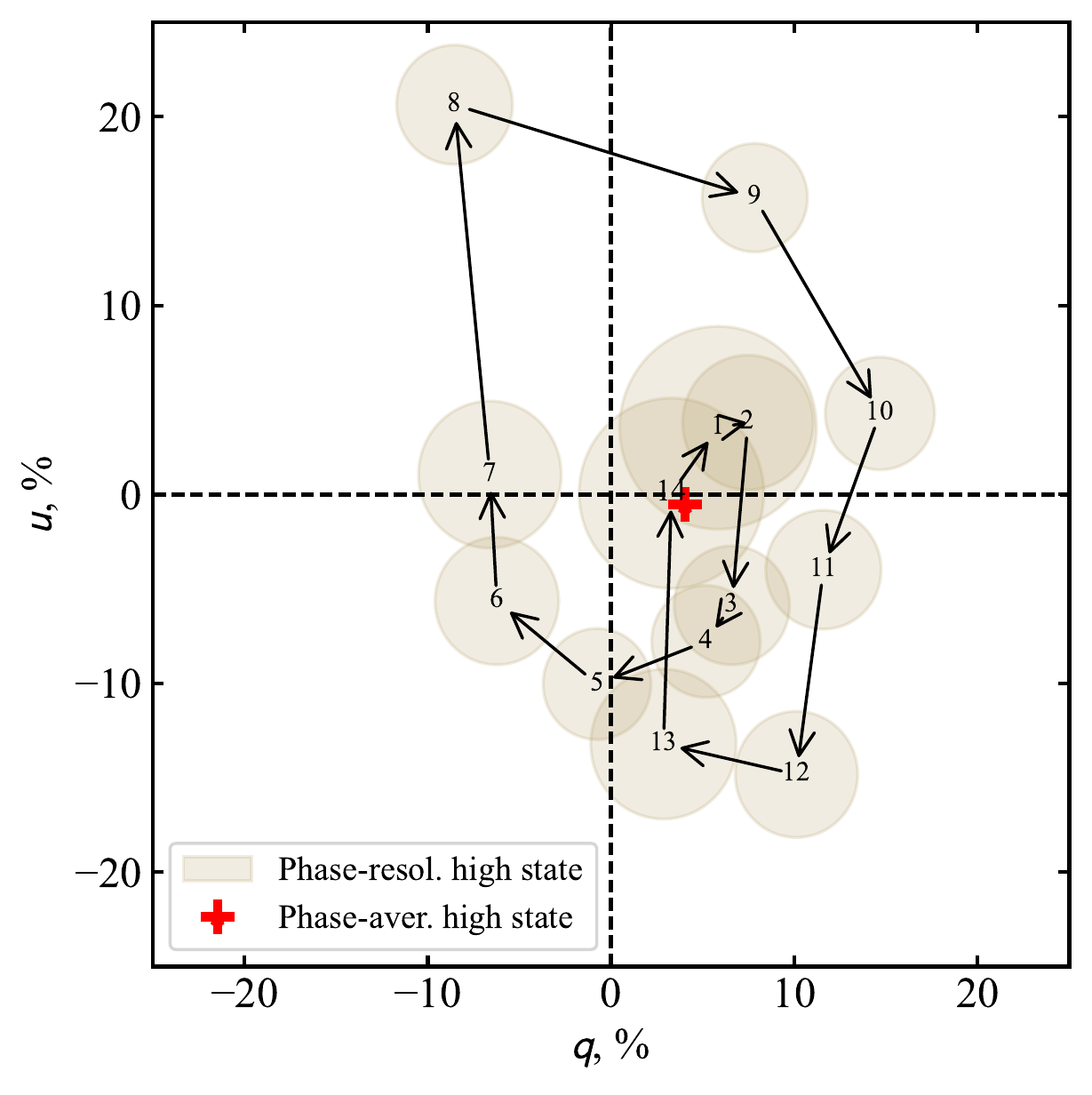}
\includegraphics[width=0.45\linewidth]{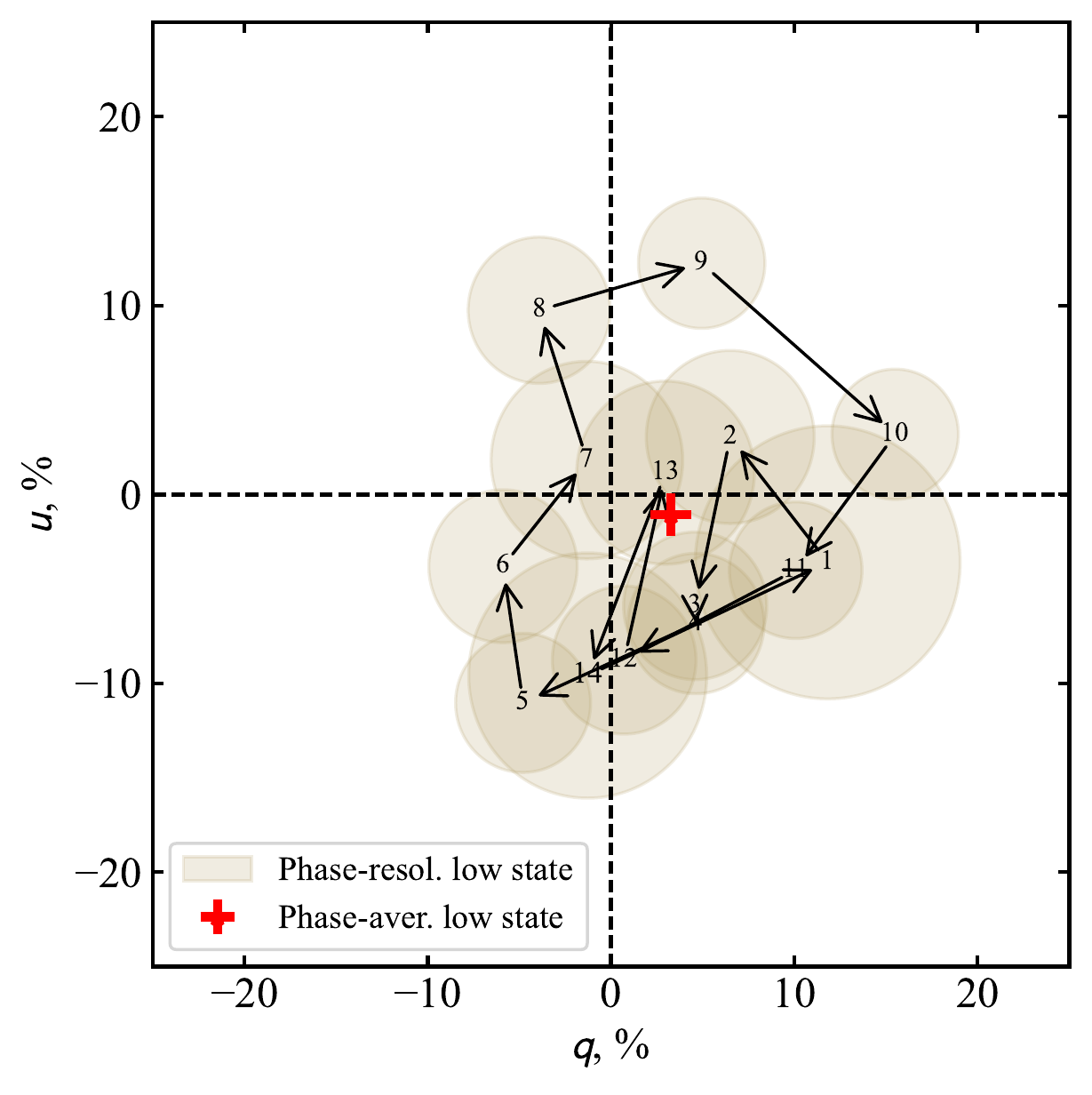}
\caption{Variations of the normalized Stokes parameters $q$ and $u$ with the pulsar phase, averaged over all DUs in the 2--8 keV energy band (beige circles and arrows) for the bright (left) and low (right) states of \gro. Each circle corresponds to a specific  phase bin, numbered successively following the binning shown in Fig.~\ref{fig:ixpe-st.pd.pa}. The circle radius represents the 1$\sigma$ uncertainty value.
The phase-averaged value is shown with a red cross in the corresponding panel.  
 }
 \label{fig:pcub-res}
\end{figure*}

The high counting statistics allowed us to measure the spin period of the NS with good accuracy, $P_{\rm spin-high}=93.133(3)$~s and $P_{\rm spin-low}=93.146(4)$~s, in the first and second \ixpe observations, respectively (the reported values and uncertainties are estimated using the phase-connection technique following \citealt{1981ApJ...247.1003D}). The pulsed fraction ---defined as $PF = (F_{\max} - F_{\min})/(F_{\max} + F_{\min})$, where  $F_{\max}$ and $F_{\min}$ are the maximum and minimum count rates in the pulse profile constructed using 80 phase bins--- was found to be around 60\% in both flux states.
The resulting pulse profiles in four energy bands are shown in Figure~\ref{fig:ixpe-pprof}. The pulse profile has a double-peaked shape with the first peak gradually disappearing at higher energies, which is peculiar for this source \citep[see, e.g.,][]{2011MNRAS.413..241N}. We see that the pulse profiles in the two luminosity states agree well, which already indicates a lack of significant changes in the accretion geometry.

\begin{figure*}
\centering
\includegraphics[width=0.4\linewidth]{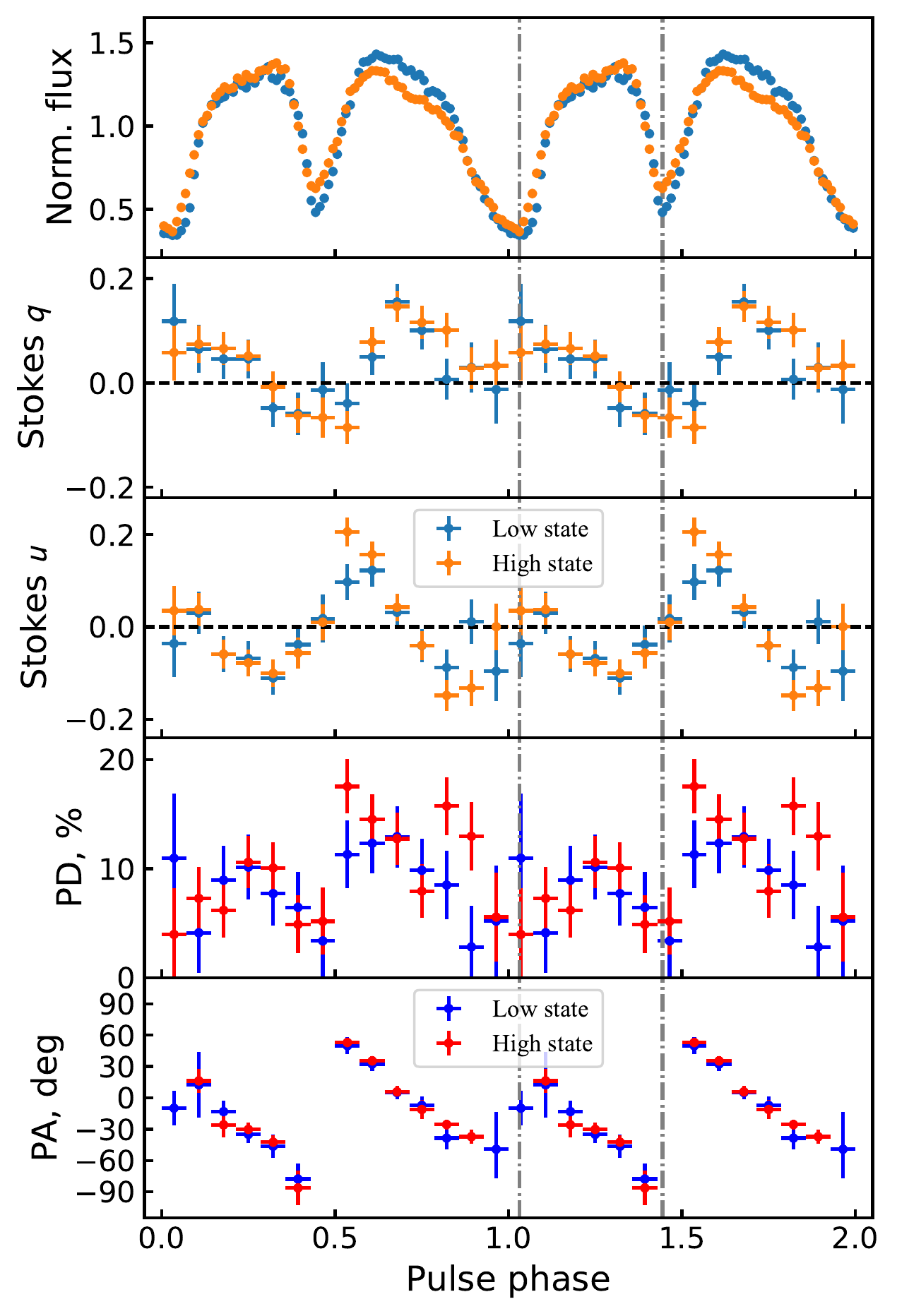}
\includegraphics[width=0.4\linewidth]{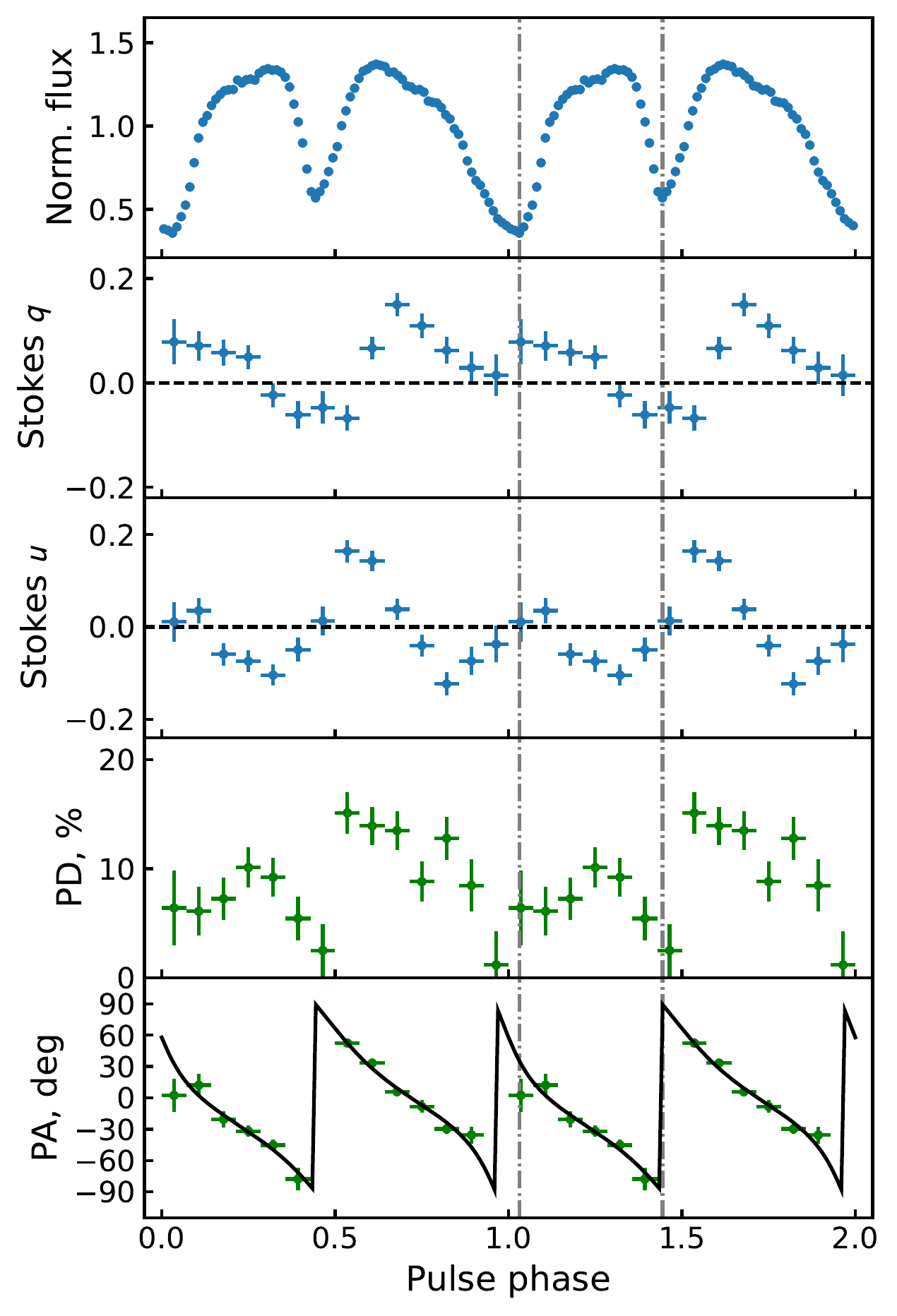}
\caption{Dependence of the normalized flux in the 2--8 keV energy band, normalized Stokes parameters $q$ and $u$ (based on the formalism by \citealt{2015APh....68...45K}), PD, and PA (from the spectro-polarimetric analysis) on the pulse phase for two \ixpe observations separately (left) and combined (right). Data from the three \ixpe telescopes are combined. Vertical dash-dotted lines show the positions of two minima in the profile and are added for illustrative purposes. The black solid line in the bottom right panel shows the best-fit rotating vector model (see Sect.~\ref{sec:geom}). 
 } 
 \label{fig:ixpe-st.pd.pa}
\end{figure*}

\subsection{Polarimetric analysis}
\label{sec:polar}
To understand whether also the polarimetric analysis supports this conclusion, we conducted pulse phase-averaged and phase-resolved analysis of \ixpe data for both observations individually.
First, we performed the polarimetric analysis of the data using the formalism by \citet{2015APh....68...45K} implemented in the {\sc ixpeobssim} package \citep{Baldini2022} under the {\tt pcube} algorithm in the {\tt xpbin} tool. Phase-averaged analysis in the 2--8 keV energy band resulted in low PD values of $4.1\pm0.9$\% and $3.4\pm1.1$\% for the bright and low states, respectively, with corresponding polarization angle (PA, measured from north to east) values of $-3\fdg8\pm6\fdg4$ and $-9\fdg1\pm9\fdg2$; that is, the results are consistent within uncertainties. However, we note that, because the PA is expected to be strongly variable over the pulsar spin phase, these phase-averaged results do not contain a significant amount of useful information. 

Therefore, as the next step, we performed a phase-resolved polarimetric analysis. For that, we used all the available data in each observation in the 2--8 keV band binned into 14 phase bins. The number of phase bins was selected to trace the strongly variable PA from one side and to get sufficiently high statistics from another. The results are shown in Figures~\ref{fig:pcub-res} and \ref{fig:ixpe-st.pd.pa}. We see that the normalized Stokes parameters $q=Q/I$ and $u=U/I$ are indeed strongly variable over the pulse phase, resulting in a relatively low phase-averaged PD value.

To properly take the energy dispersion and the spectral shape  into account, we also conducted a spectro-polarimetric analysis through a joint fit of the $I$, $Q,$ and $U$ spectra (prepared with the PHA1, PHA1Q, and PHA1U algorithms in the {\tt xpbin} tool) using the {\sc xspec} package \citep[see][]{2017ApJ...838...72S}. 
Although the broadband spectrum of the source depends on the luminosity, below 10\,keV its shape can be described with a simple absorbed power law in a very broad range of luminosities \citep{2017A&A...608A..17T}. To account for the remaining calibration uncertainties in the \ixpe data, as well as possible spectral complexity due to variations of the spectral parameters with phase,  we used a more flexible model of a power law with the high-energy exponential rolloff ({\tt cutoffpl} in {\sc xspec}) to model the phase-averaged spectra.

\begin{table}
    \caption{Spectral parameters for the best-fit model for the two flux states of the source, separately and combined; uncertainties are at 68.3\% CL.}
    \label{tab:spec-aver}
    \centering
    \begin{tabular}{rll}
    \hline\hline
    Parameter & Value & Units\\ \hline
    \multicolumn{3}{c}{Bright state}  \\
    $N_{\rm H}$  & 0.7$\pm0.2$ &  $10^{22}$~cm$^{-2}$ \\
    const$_{\rm DU2}$ & 0.964$\pm0.004$ & \\
    const$_{\rm DU3}$ & 0.924$\pm0.004$ & \\
    Photon index & $-0.7\pm0.1$ & \\
    E-folding energy & $3.2\pm0.3$ & keV \\
    PD & 3.8$\pm0.7$ & \% \\
    PA & $-$5.8$\pm5.3$ & deg \\
    Flux (2--8~keV) & 10.31$\pm0.05$& $10^{-10}$ erg~cm$^{-2}$~s$^{-1}$ \\
    Luminosity (2--8~keV) & $1.6\times10^{36}$  & erg~s$^{-1}$ at $d=3.6$~kpc\\ 
    $\chi^{2}$ (d.o.f.) & 1391 (1333) & \\
    \hline
    \multicolumn{3}{c}{Low state}  \\
    $N_{\rm H}$  & 0.8$\pm0.2$ &  $10^{22}$~cm$^{-2}$ \\
    const$_{\rm DU2}$ & 0.962$\pm0.005$ & \\
    const$_{\rm DU3}$ & 0.928$\pm0.005$ & \\
    Photon index & $-0.8\pm0.2$ & \\
    E-folding energy & $2.7\pm0.2$ & keV \\
    PD & 3.9$\pm0.9$ & \% \\
    PA & $-$6.7$\pm6.4$ & deg \\
    Flux (2--8~keV) & 5.46$\pm0.03$& $10^{-10}$ erg~cm$^{-2}$~s$^{-1}$ \\
    Luminosity (2--8~keV) & $0.9\times10^{36}$  & erg~s$^{-1}$ at $d=3.6$~kpc\\ 
    $\chi^{2}$ (d.o.f.) & 1371 (1297) & \\
    \hline
    \multicolumn{3}{c}{Combined data}  \\
    $N_{\rm H}$ (bright)   & 0.7$\pm0.1$ &  $10^{22}$~cm$^{-2}$ \\
    $N_{\rm H}$ (low)      & 0.8$\pm0.2$ &  $10^{22}$~cm$^{-2}$ \\
    Photon index (bright) & $-0.7\pm0.1$ & \\
    E-folding energy (bright) & $3.2\pm0.3$ & keV \\
    Photon index (low)   & $-0.8\pm0.2$ & \\
    E-folding energy (low) & $2.7\pm0.2$ & keV \\
    PD & 3.9$\pm0.5$ & \% \\
    PA & $-$6.2$\pm4.1$ & deg \\
    $\chi^{2}$ (d.o.f.) & 2762 (2632) & \\
    \hline
    \end{tabular}
\end{table}

We first applied the model consisting of a rolloff power law modified by the interstellar absorption \citep[{\tt tbabs} in {\sc xspec} with abundances adopted from][]{Wilms2000} to the phase-averaged data in the bright and low states, separately. The rolloff power-law  component was combined with a constant polarization model (energy-independent PD and PA), {\tt polconst} in {\sc xspec}. An additional multiplicative constant {\tt const} was introduced to account for possible discrepancies in absolute effective area calibration of independent DUs; the constant was fixed to unity for DU1, taken as a reference. The final form of the model in {\sc xspec} is {\tt const$\times$tbabs$\times$polconst$\times$cutoffpl}.

\begin{figure*}
\centering
\includegraphics[width=0.313\linewidth]{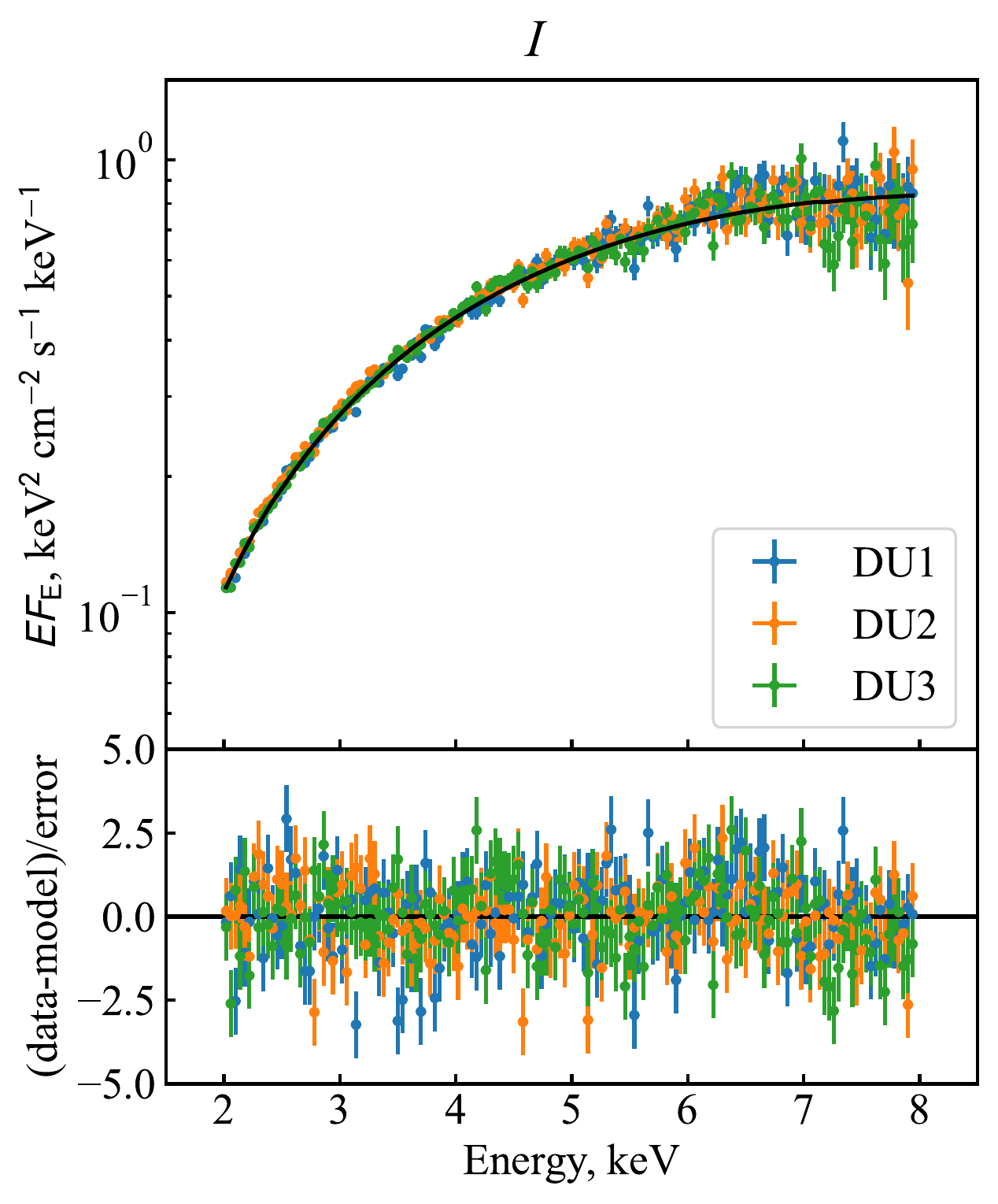}
\includegraphics[width=0.32\linewidth]{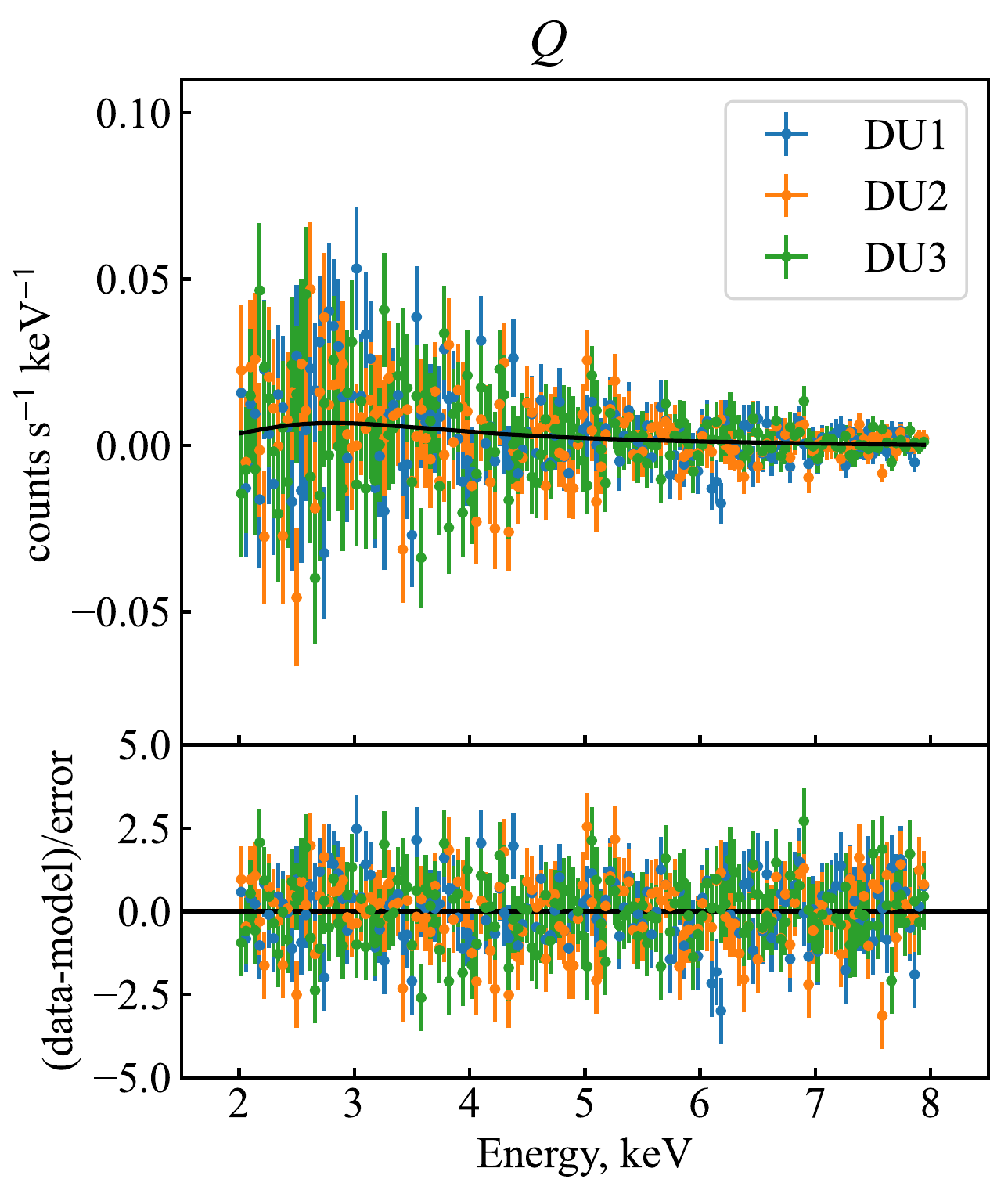}
\includegraphics[width=0.32\linewidth]{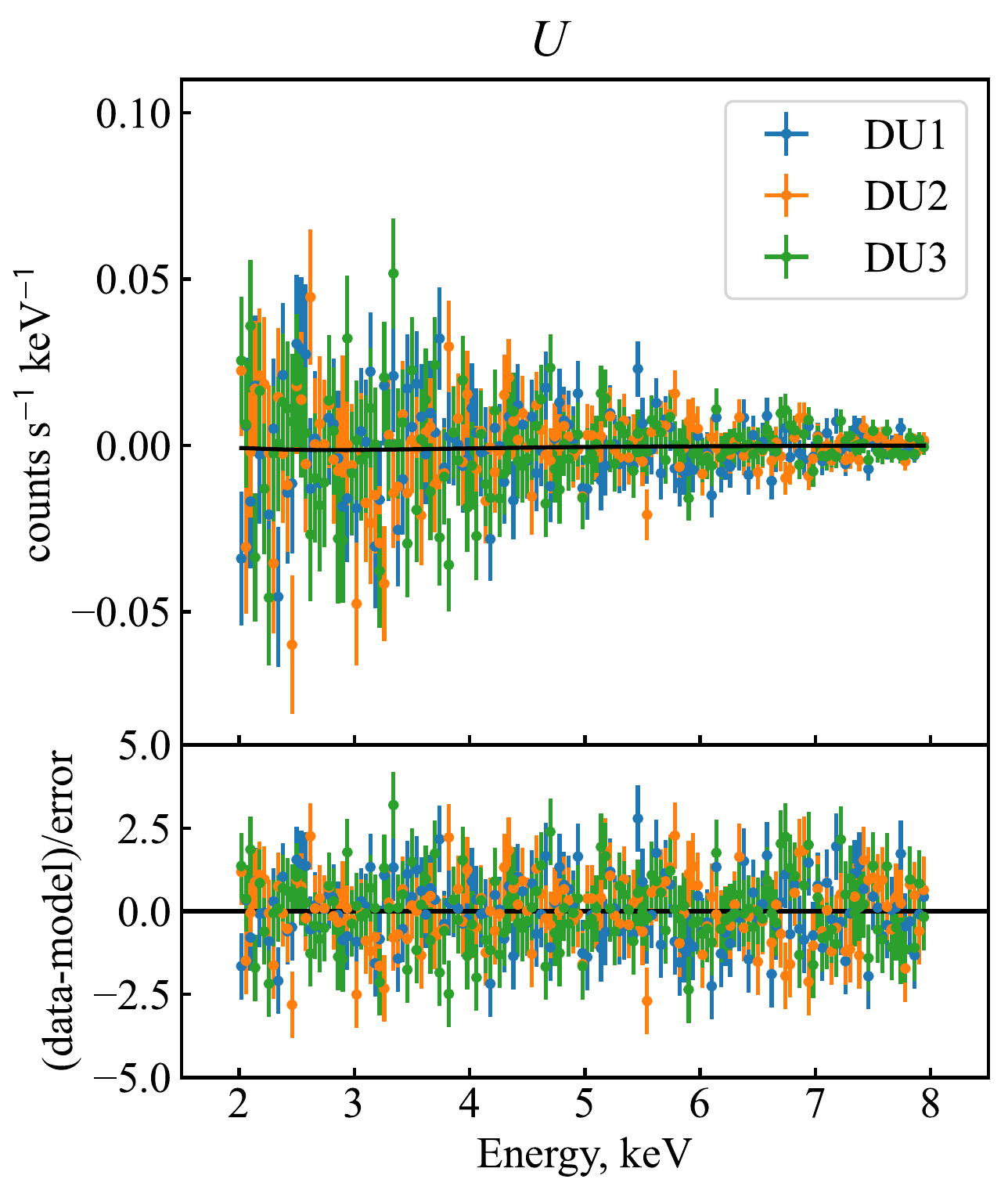}

\includegraphics[width=0.313\linewidth]{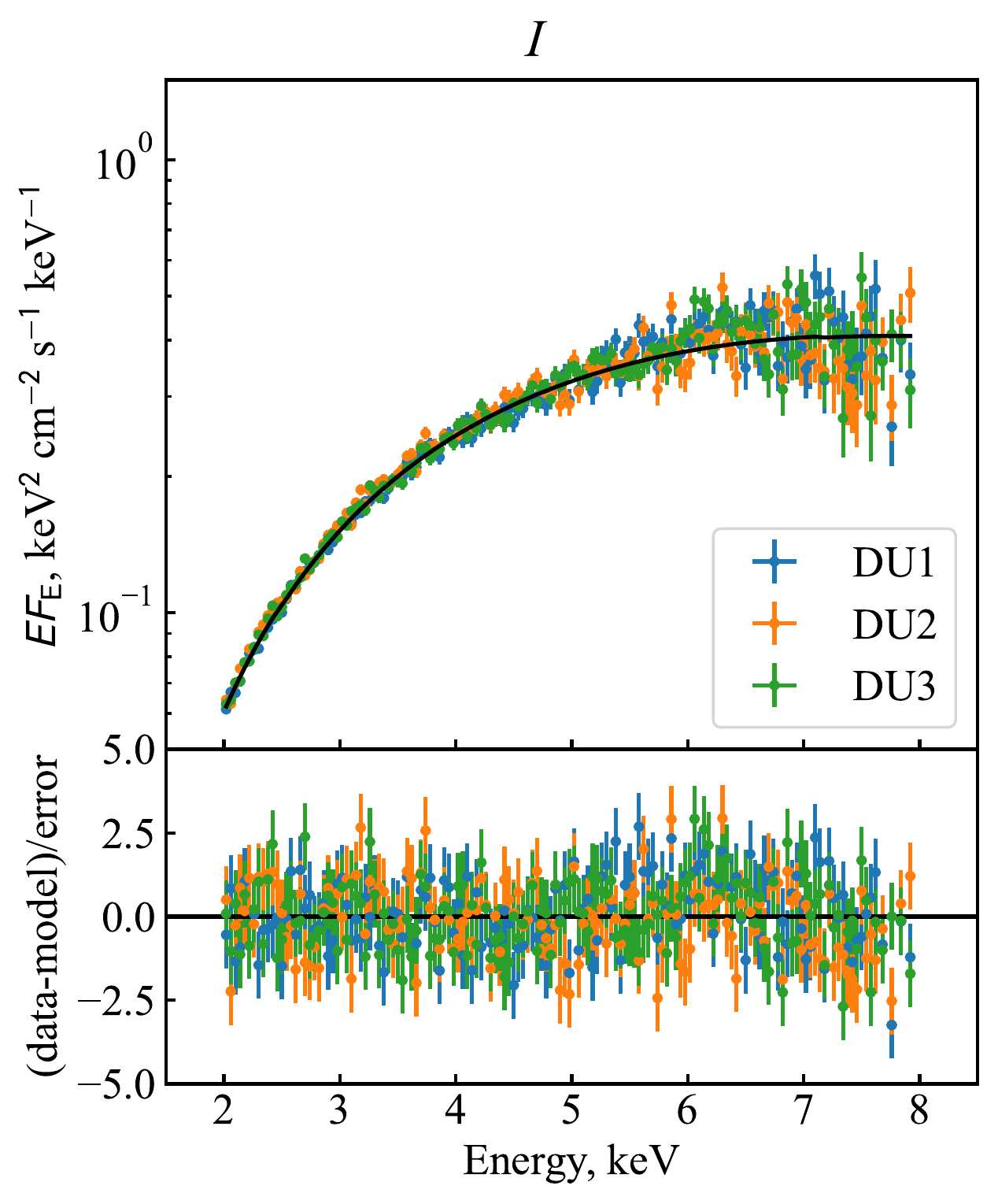}
\includegraphics[width=0.32\linewidth]{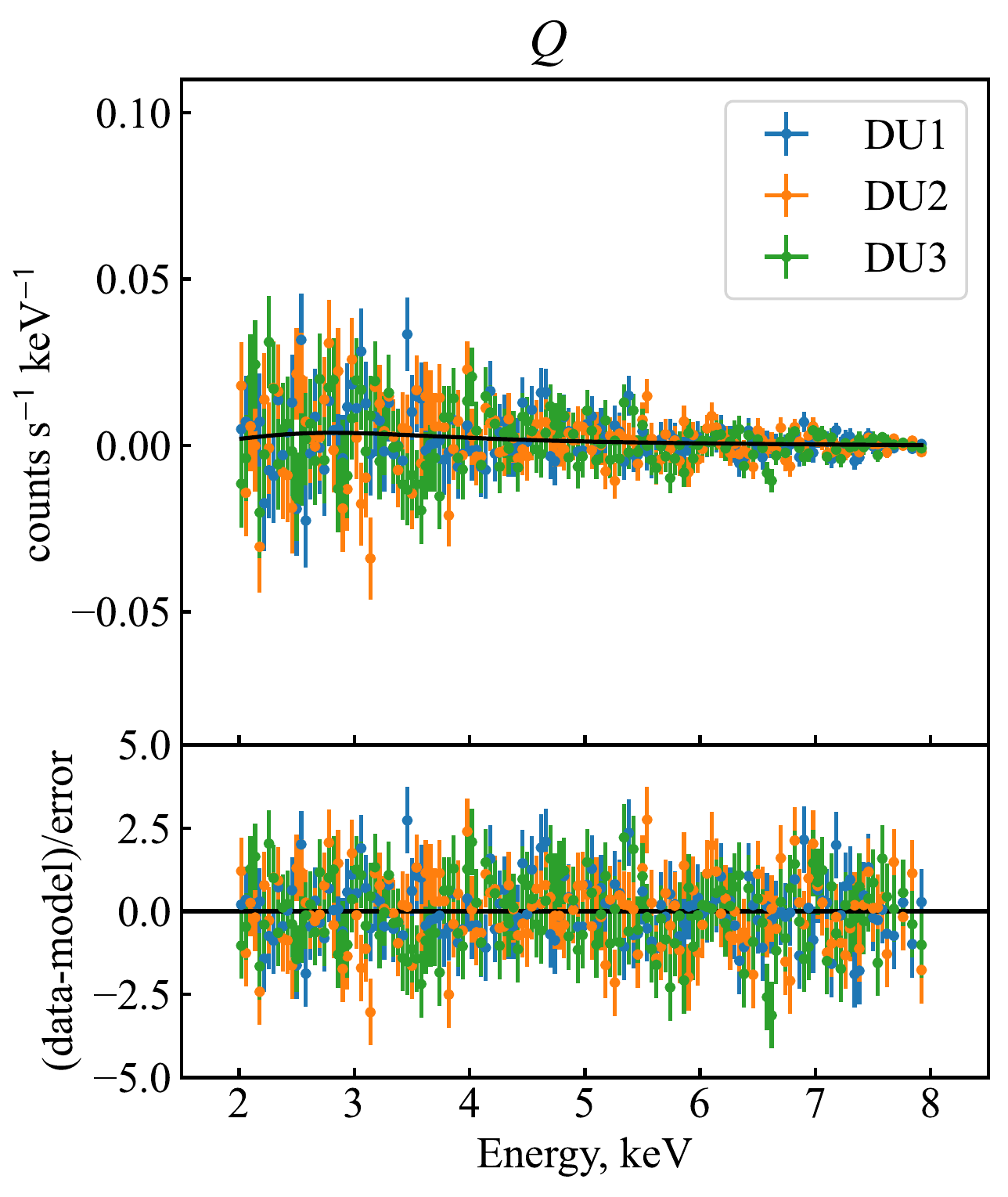}
\includegraphics[width=0.32\linewidth]{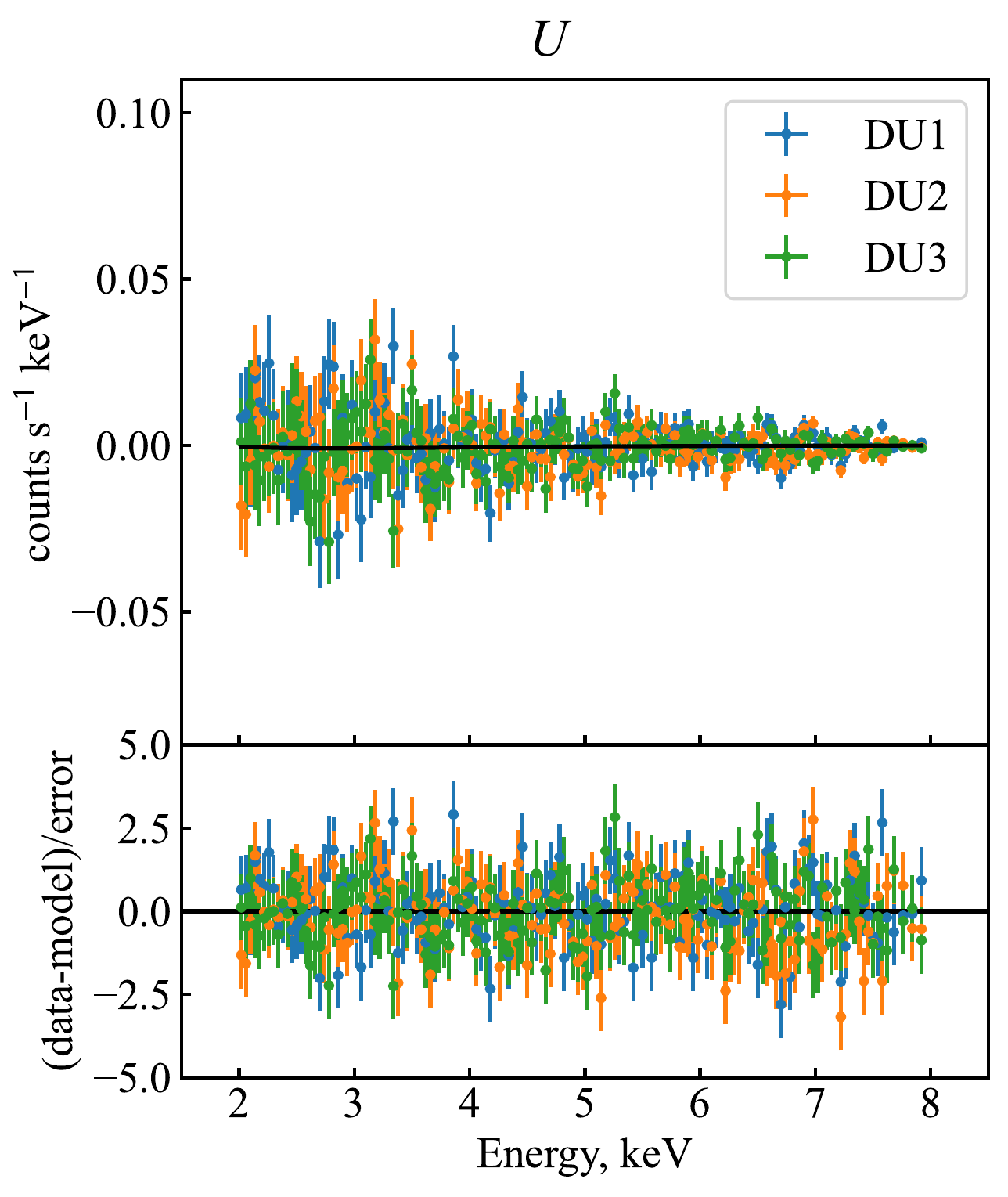}

\caption{Energy distributions of the Stokes parameters $I$, $Q,$ and $U$ for the bright (upper plots) and low (lower plots) states of \gro with the best-fit model shown with the black solid line. The residuals between the data and the model normalized for the errors are shown in the bottom panels of the corresponding plots. The different colors represent the three \ixpe detectors: DU1 in blue, DU2 in orange, and DU3 in green. }
 \label{fig:specs}
\end{figure*}

As a result, we obtained polarization parameters fully consistent with values obtained from the energy-binned analysis using the {\tt pcube} algorithm. The parameters of the best-fit model are presented in Table~\ref{tab:spec-aver}. The quality of the obtained fits for both observations can be seen from Fig.~\ref{fig:specs}, where the energy spectra for $I$, $Q,$ and $U$ Stokes parameters are shown with the corresponding residuals. 

To study the energy spectrum and polarization properties of \gro as a function of the spin phase, we performed the phase-resolved spectro-polarimetric analysis using the same 14 phase bins as above. For the spectral fit, we used a simpler continuum model consisting of a power law ({\tt const$\times$tbabs$\times$powerlaw$\times$polconst}) with the cross-calibration constants fixed at the values derived from the phase-averaged analysis (see Table~\ref{tab:spec-aver}). The fit results (the PD and PA values) are presented in Fig.~\ref{fig:ixpe-st.pd.pa}. 

\begin{figure*}
\centering
\includegraphics[width=0.3\linewidth]{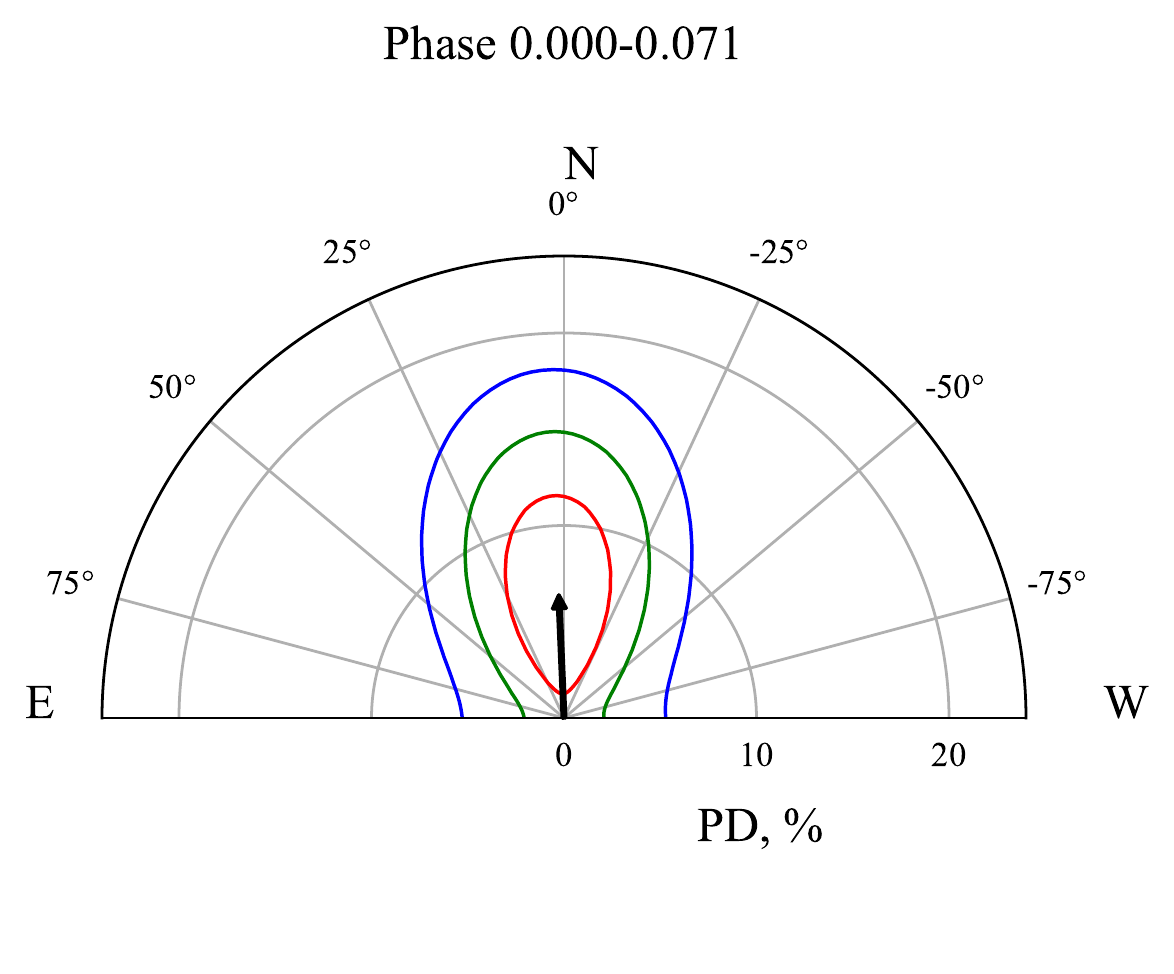}
\includegraphics[width=0.3\linewidth]{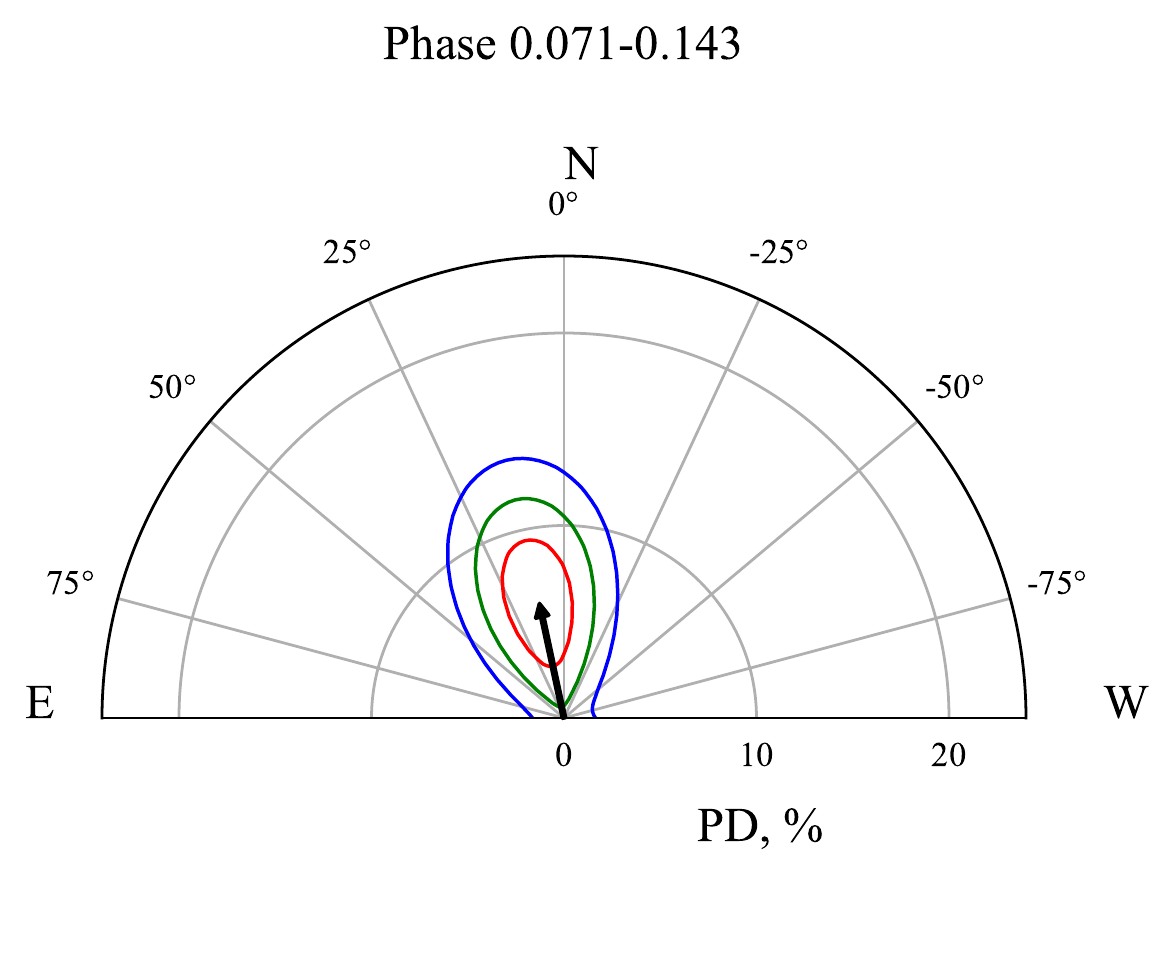}
\includegraphics[width=0.3\linewidth]{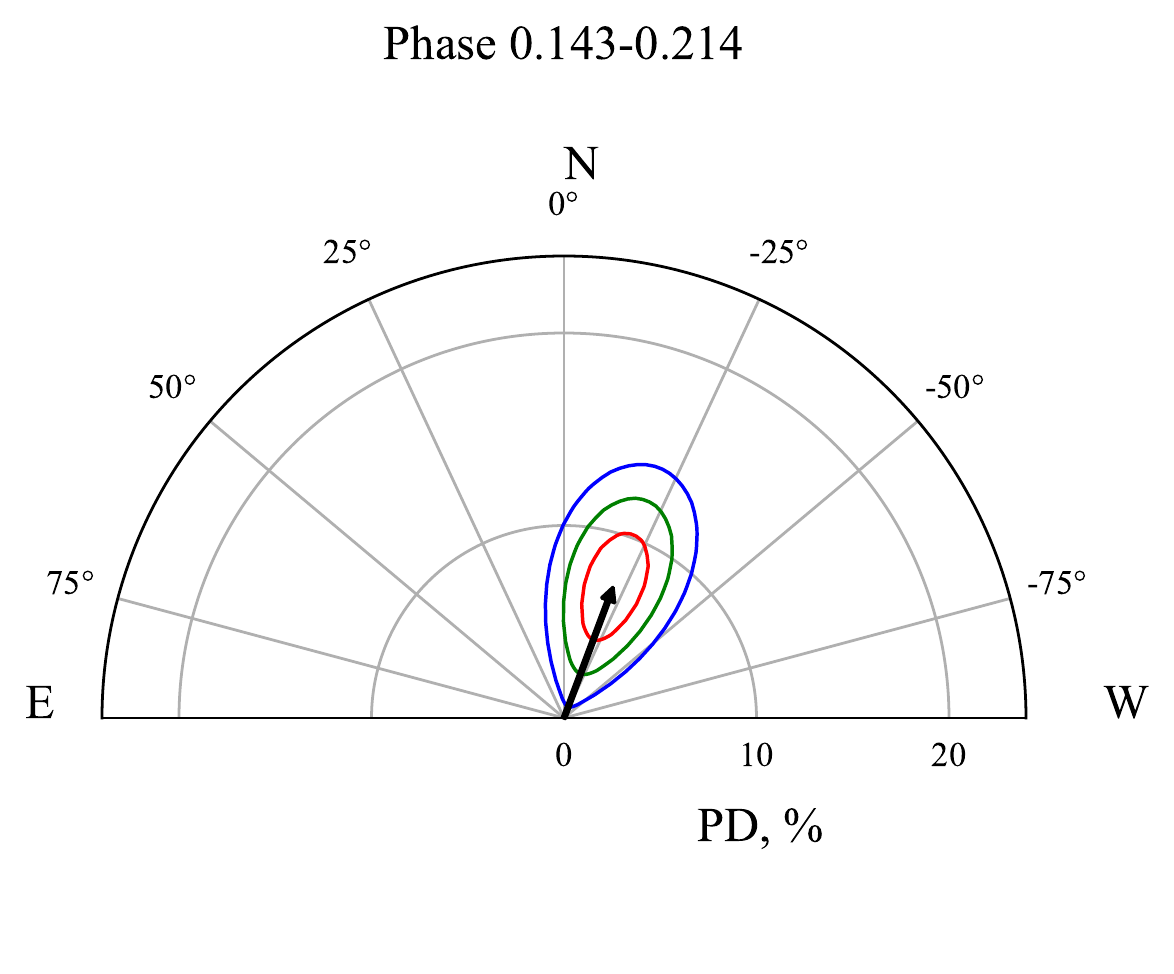}
\includegraphics[width=0.3\linewidth]{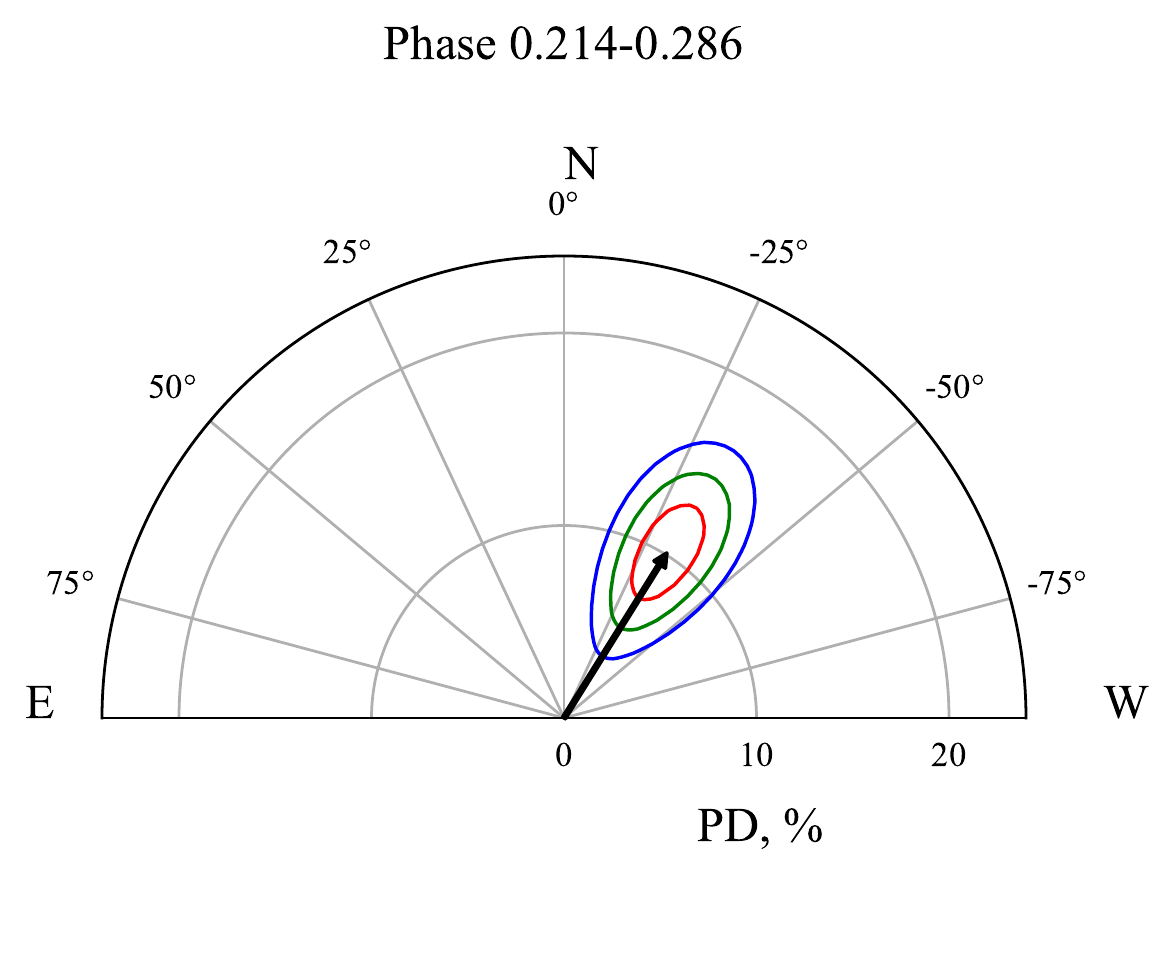}
\includegraphics[width=0.3\linewidth]{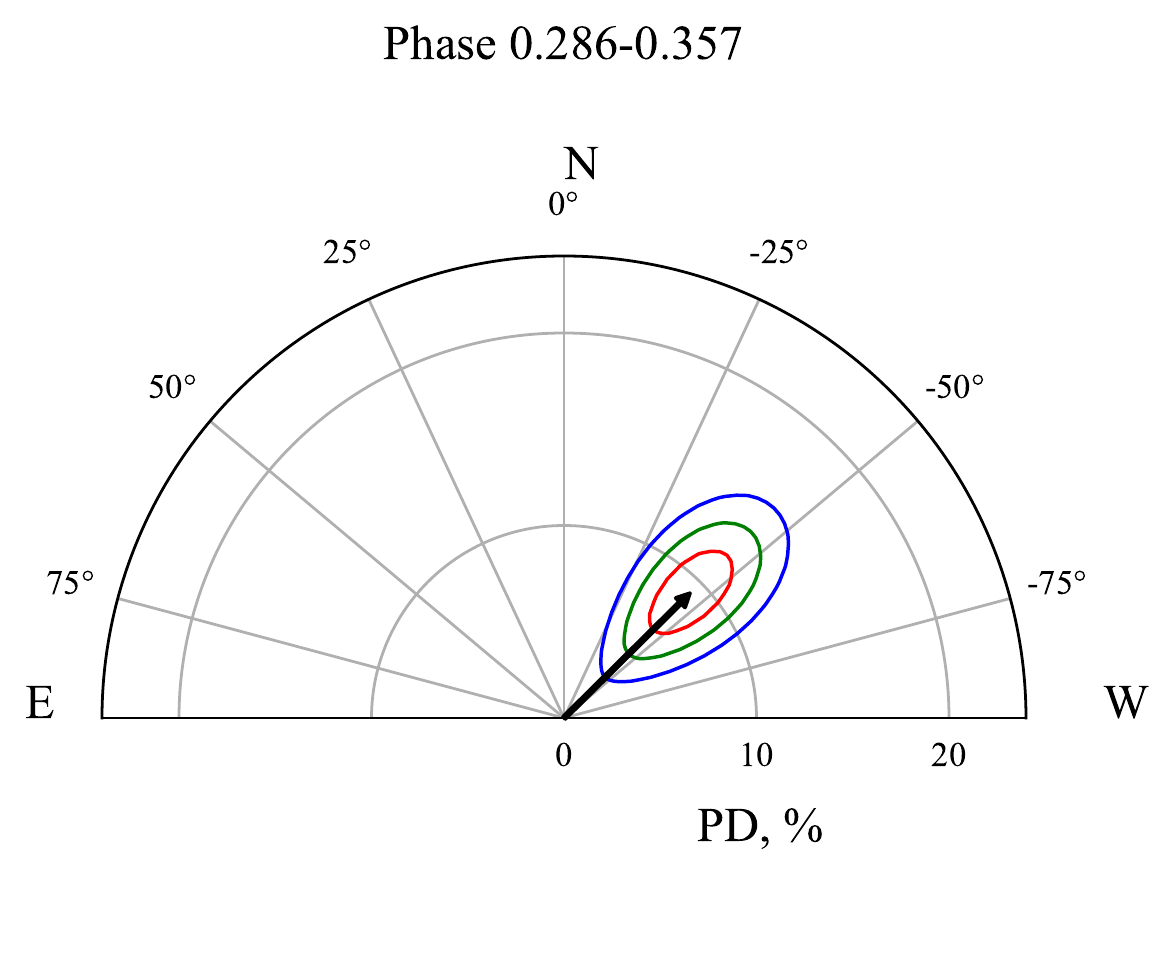}
\includegraphics[width=0.3\linewidth]{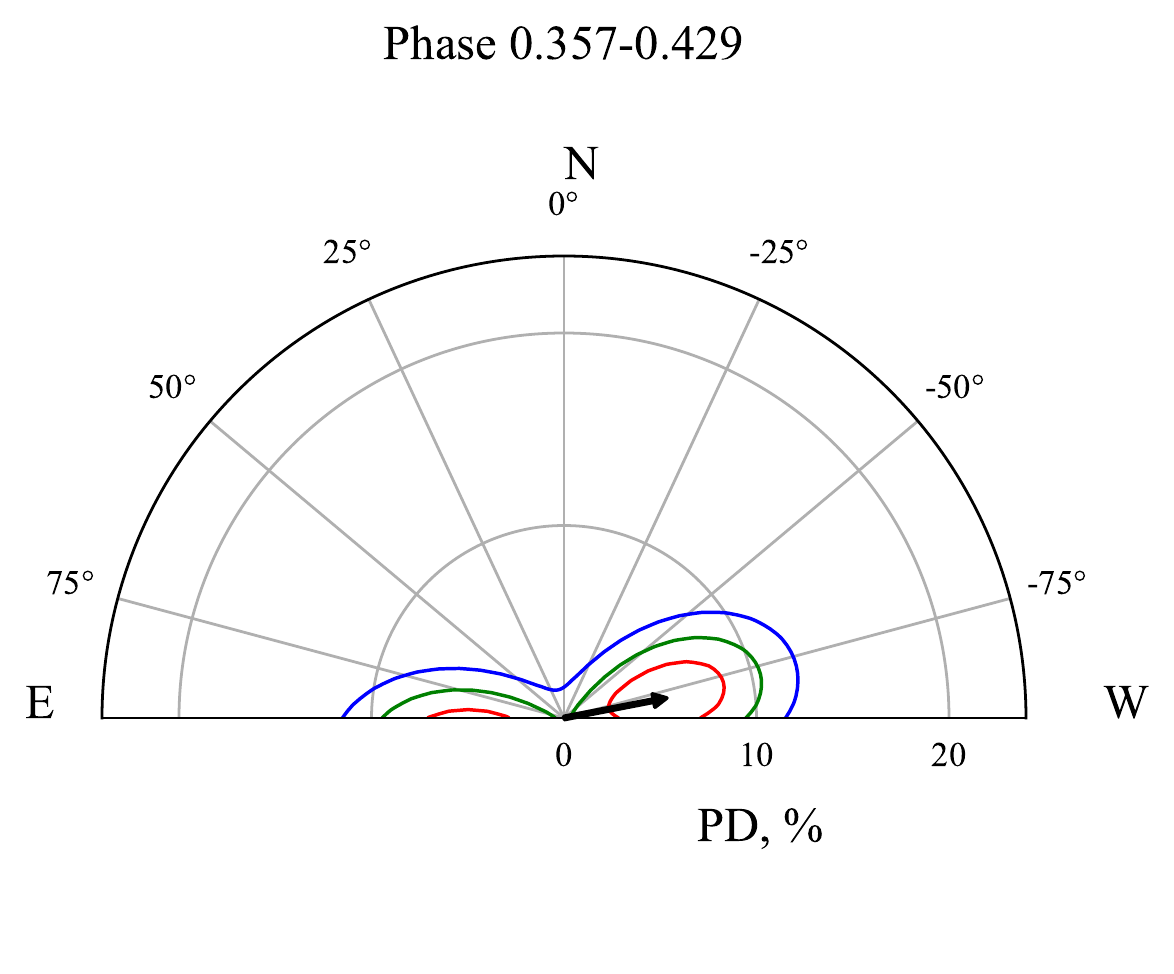}
\includegraphics[width=0.3\linewidth]{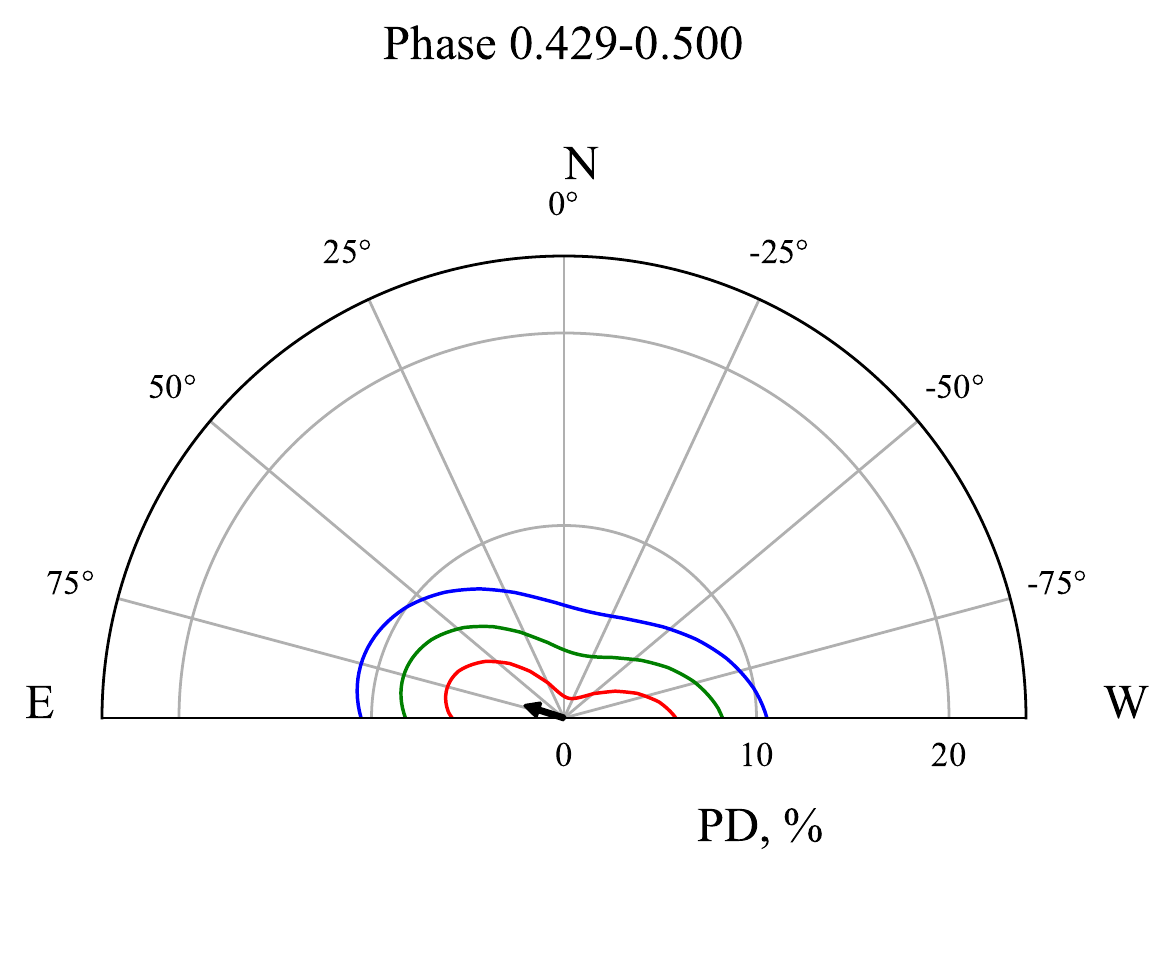}
\includegraphics[width=0.3\linewidth]{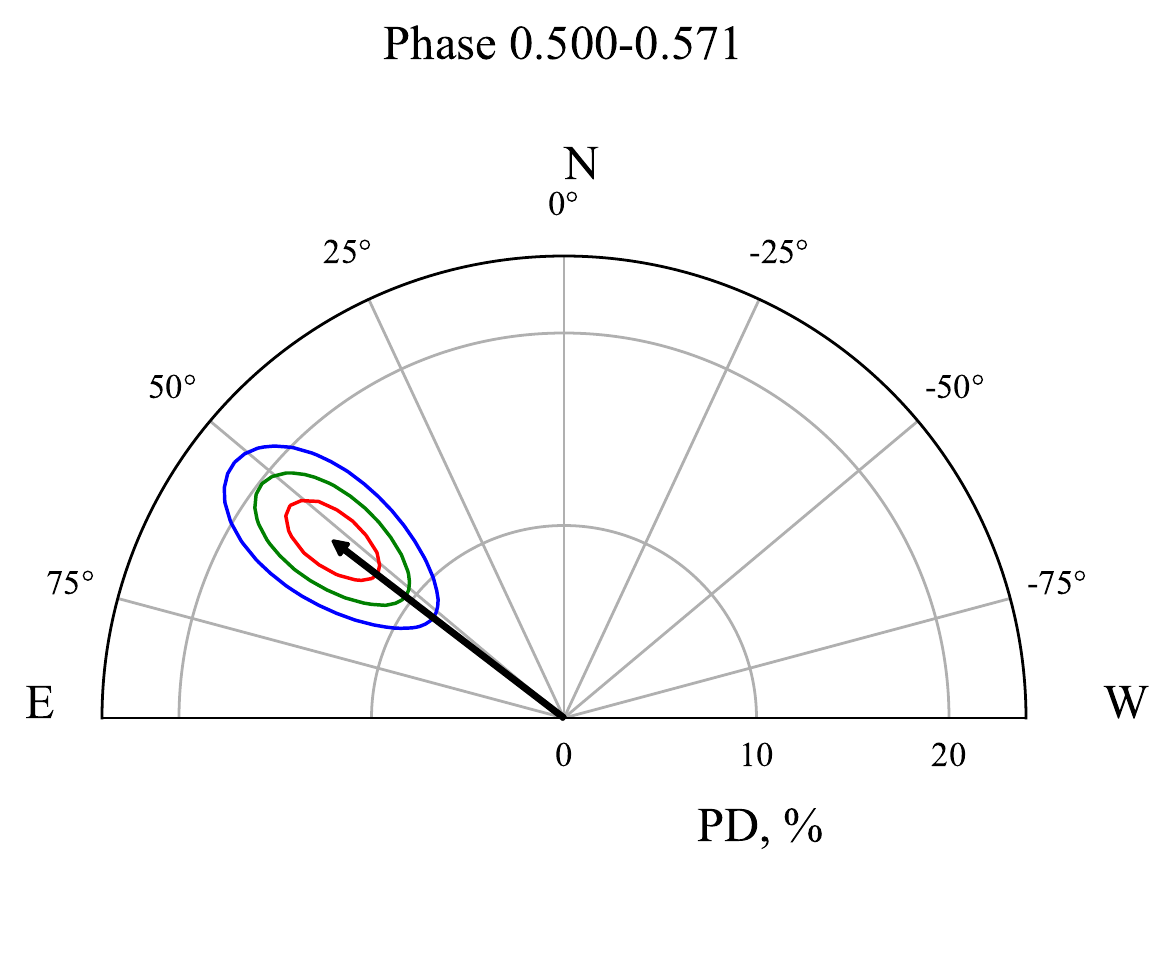}
\includegraphics[width=0.3\linewidth]{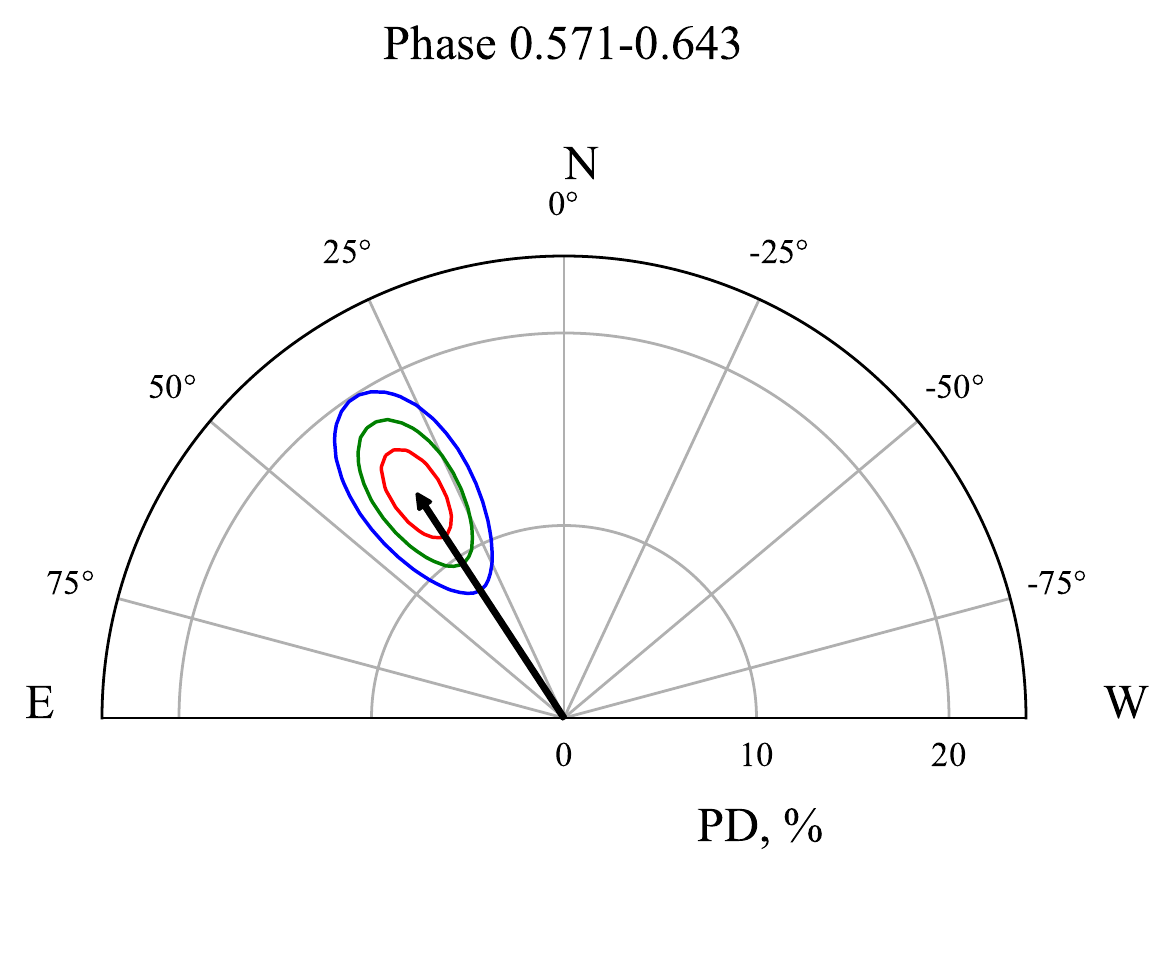}
\includegraphics[width=0.3\linewidth]{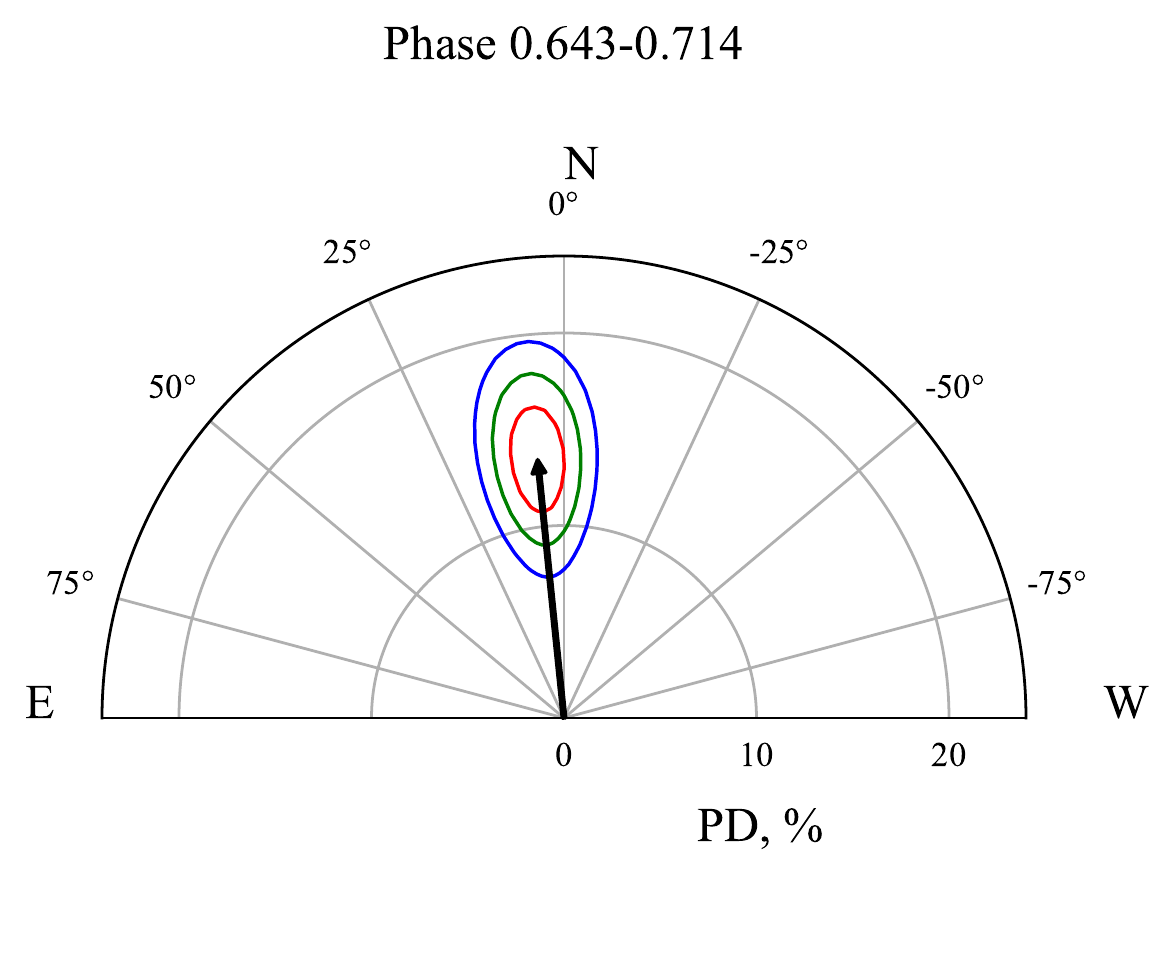}
\includegraphics[width=0.3\linewidth]{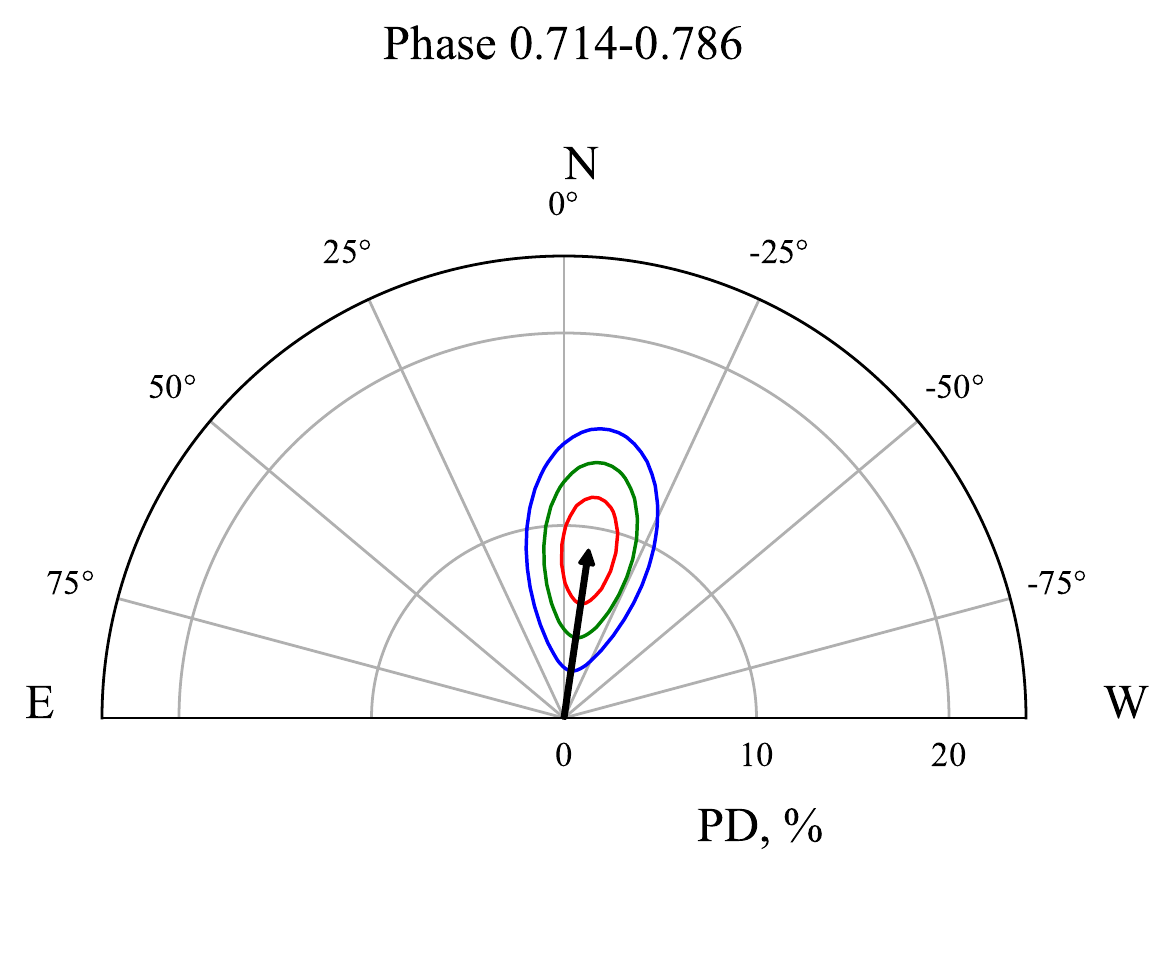}
\includegraphics[width=0.3\linewidth]{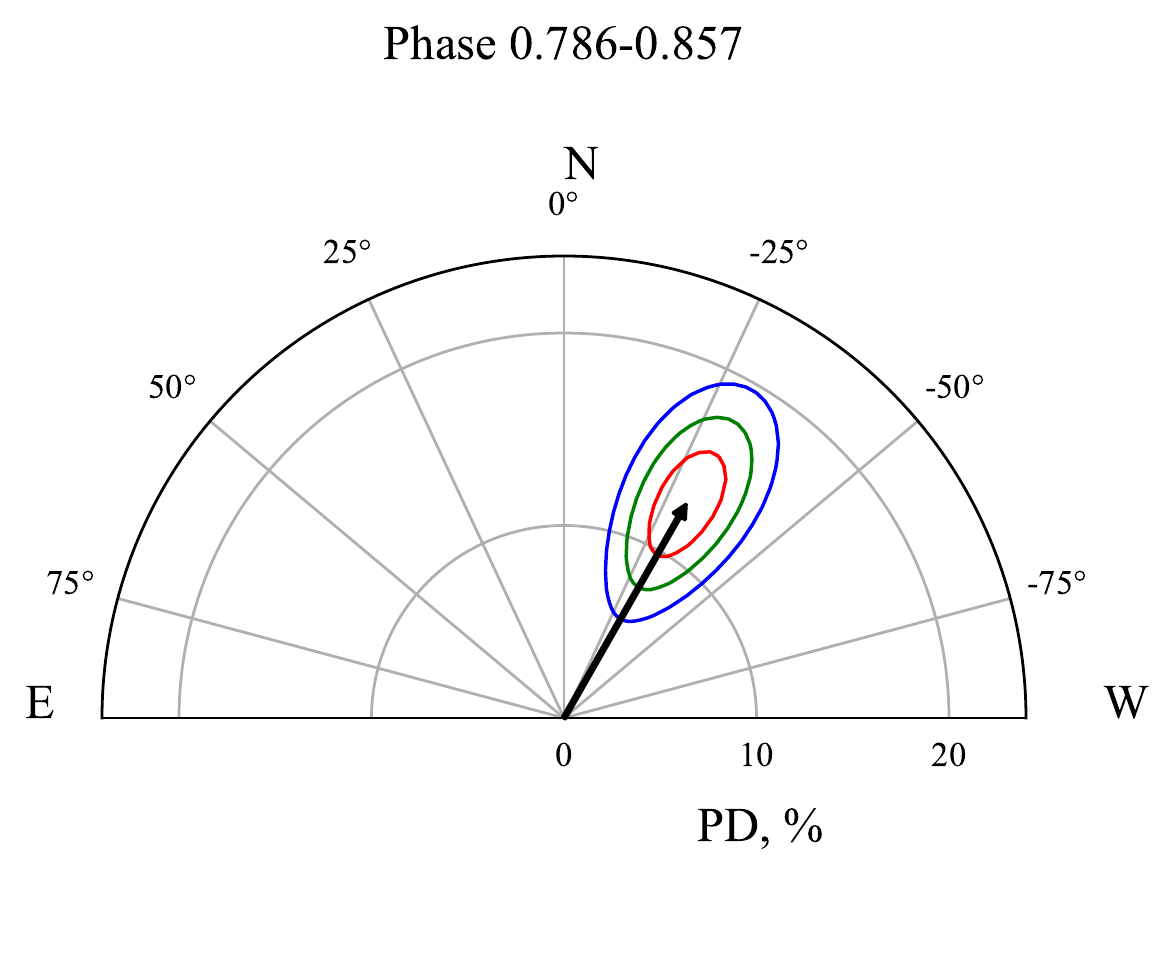}
\includegraphics[width=0.3\linewidth]{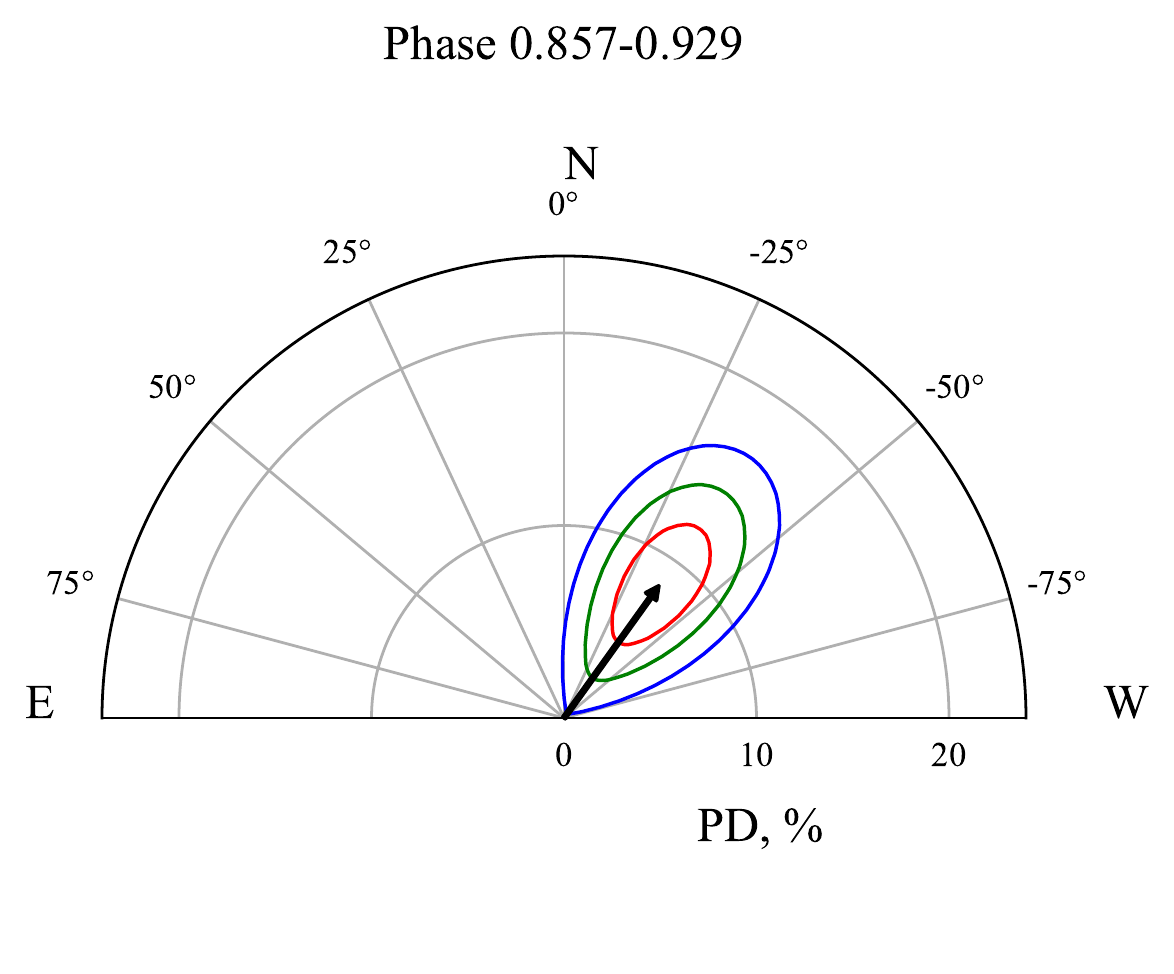}
\includegraphics[width=0.3\linewidth]{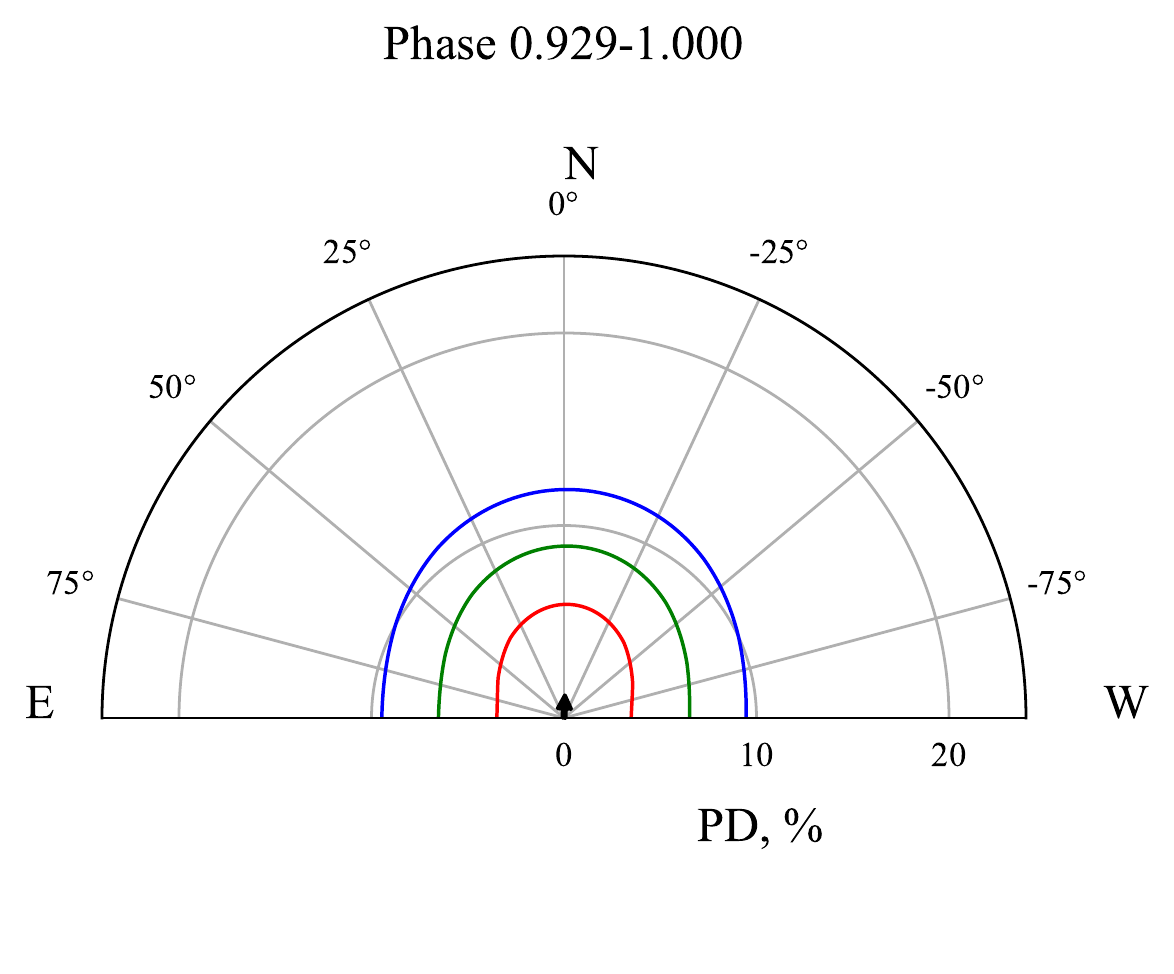}
\caption{Polarization vectors as a function of the phase of \gro based on the spectral fitting of the combined data from both \ixpe observations.  In each plot, the PD and PA contours at 68.27\%, 95.45\%, and 99.73\% confidence levels (red, green and blue, respectively) are shown in polar coordinates for 14 different phase intervals (coinciding with the ones defined in Fig.~\ref{fig:ixpe-st.pd.pa}).}
 \label{fig:cont-resol}
\end{figure*}

We see from Fig.~\ref{fig:ixpe-st.pd.pa} that there is no significant difference in the polarization properties of the source during different observations. Therefore, to increase the counting statistics, we combined the data from both observations after the phase alignment using the broadband pulse profiles. After that, we repeated both phase-averaged and phase-resolved spectro-polarimetric analysis applying the same phase binning and spectral model. To take into account the dependence of the spectral shape on the \gro luminosity in the joint fit, we allowed the absorption value $N_{\rm H}$, the photon index, and the normalization to vary independently in the two data sets, whereas PD and PA were tied. In the phase-resolved analysis, all these parameters (including $N_{\rm H}$) were allowed to vary over the spin phase to reflect possible inhomogeneities in the flow of matter around the NS. An energy-binned ({\tt pcube}) analysis was performed on the combined dataset as well.
The obtained results are summarized in Figs.~\ref{fig:ixpe-st.pd.pa} and \ref{fig:cont-resol} and Tables~\ref{tab:spec-aver} and \ref{tab:fit_phbin}.

Modification of the {\tt polconst} polarization component of the best-fit model with {\tt pollin} and {\tt polpow}, which correspond to a linear and a power-law dependence of PD and PA on energy, respectively, did not lead to any significant improvement of the fit quality. Namely, for the phase-averaged data combined over both observations, this modification resulted in $\Delta \chi^2\sim1$ for 2 d.o.f. for both {\tt pollin} and {\tt polpow} models. For the phase-resolved data, and for
each phase independently, the $\Delta \chi^2$ value did not exceed $\sim5$ for 2 d.o.f. with a corresponding F-test probability of $\sim0.1$. To increase the counting statistics in order to reveal the energy dependence of PD, we applied the method proposed by \cite{Mushtukov23}. Namely, we simultaneously fitted spectra from all 14 phase bins using the {\tt pollin} model with the PAs frozen at the predictions of the geometrical model discussed in Sect.~\ref{sec:geom}. 
Photon indices were fixed at the values presented in Table~\ref{tab:fit_phbin} and the PD parameters (i.e., normalization at 1 keV, $A_1$, and the slope of energy dependence, $A_{\rm slope}$) were tied for all phases in both observations. 
As a result, we obtained a phase-averaged PD at 1 keV of $A_1=3.5\%\pm1.8\%$ and a slope of $A_{\rm slope}=2.1\%\pm0.6\%$~keV$^{-1}$. The significance of the improvement of the fit with the \texttt{pollin} model compared to \texttt{polconst} based on the F-test is $2\times10^{-4}$. This can be considered as a strong indication that the PD of \gro depends on energy. A similar conclusion was recently reached by \citet{Forsblom2023}  and \citet{Mushtukov23} for Vela~X-1 and X Persei, respectively.

\begin{table*}
    \caption{Spectral parameters for the phase-resolved spectro-polarimetric analysis of the combined data. }
    \centering
    \begin{tabular}{cccccccc}
    \hline\hline
       Phase  &   $N_{\rm H}$ (bright) &   $N_{\rm H}$ (low) &  Photon index &  Photon index &  PD & PA   & $\chi^{2}$ (d.o.f.) \\ 
          &    (10$^{22}$ cm$^{-2}$) &    (10$^{22}$ cm$^{-2}$) &  (bright)  &  (low)  &  (\%) & (deg)  &   \\
          \hline
0.000--0.071 & 0.9$\pm0.3$  &  0.9$\pm0.5$  &  0.87$\pm0.09$  &  1.06$\pm0.13$   &   6.4$\pm3.4$  &   2.4$\pm16.0$   &  1133 (1123) \\
0.071--0.143 & 3.2$\pm0.3$  &  3.5$\pm0.3$  &  1.29$\pm0.06$  &  1.39$\pm0.08$   &   6.1$\pm2.2$  &   12.1$\pm10.7$   &  1503 (1570) \\
0.143--0.214 & 3.0$\pm0.2$  &  3.7$\pm0.3$  &  1.29$\pm0.05$  &  1.51$\pm0.07$   &   7.3$\pm1.9$  &  $-$20.6$\pm7.8$    &  1632 (1711) \\
0.214--0.286 & 3.4$\pm0.2$  &  3.4$\pm0.3$  &  1.31$\pm0.05$  &  1.38$\pm0.06$   &  10.1$\pm1.8$  &  $-$31.9$\pm5.3$   &  1728 (1764) \\
0.286--0.357 & 3.4$\pm0.2$  &  3.4$\pm0.3$  &  1.27$\pm0.05$  &  1.27$\pm0.06$   &   9.2$\pm1.8$  &  $-$45.3$\pm5.7$   &  1838 (1793) \\
0.357--0.429 & 2.5$\pm0.2$  &  3.4$\pm0.3$  &  0.81$\pm0.05$  &  1.03$\pm0.07$   &   5.4$\pm2.0$  &  $-$77.9$\pm10.9$   &  1754 (1741) \\
0.429--0.500 & 0.4$\pm0.3$  &  1.3$\pm0.4$  &  0.01$\pm0.06$  &  $-$0.07$\pm0.08$    &   2.5$\pm2.4$  &  \dots   &  1672 (1607) \\
0.500--0.571 & 3.6$\pm0.2$  &  4.1$\pm0.3$  &  0.82$\pm0.05$  &  0.91$\pm0.06$   &  15.1$\pm1.9$  &  52.5$\pm3.6$      &  1821 (1826) \\
0.571--0.643 & 3.6$\pm0.2$  &  3.7$\pm0.3$  &  0.97$\pm0.05$  &  1.13$\pm0.06$   &  13.9$\pm1.7$  &   33.2$\pm3.6$      &  1910 (1871) \\
0.643--0.714 & 3.8$\pm0.2$  &  4.1$\pm0.3$  &  1.02$\pm0.05$  &  1.21$\pm0.06$   &  13.5$\pm1.8$  &   5.8$\pm3.8$   &  1957 (1848) \\
0.714--0.786 & 3.6$\pm0.2$  &  3.6$\pm0.3$  &  0.98$\pm0.05$  &  1.10$\pm0.06$   &   8.8$\pm1.8$  &   $-$8.4$\pm6.1$   &  1935 (1841) \\
0.786--0.857 & 3.0$\pm0.2$  &  4.0$\pm0.3$  &  0.93$\pm0.06$  &  1.22$\pm0.07$   &  12.8$\pm2.0$  &  $-$29.7$\pm4.5$    &  1714 (1753) \\
0.857--0.929 & 2.6$\pm0.3$  &  3.1$\pm0.3$  &  1.05$\pm0.07$  &  1.05$\pm0.08$   &   8.5$\pm2.4$  &  $-$35.7$\pm8.1$   &  1522 (1533) \\
0.929--1.000 & 1.8$\pm0.3$  &  1.3$\pm0.4$  &  1.14$\pm0.09$  &  0.95$\pm0.11$   &   1.2$^{+2.1}_{-1.2}$  & \dots    &  1309 (1206) \\
    \hline
    \end{tabular}
   \label{tab:fit_phbin}
\end{table*}

\section{Discussion} 
\label{sec:discussion}

\subsection{Polarization mechanisms}
\label{sec:mech}

Under the condition of a strong magnetic field, the medium experiences birefringence when the phase velocity of photons depends on their polarization state.
In this case, the photons tend to propagate in the form of two orthogonal polarization modes: the ordinary (O-mode) and extraordinary (X-mode) ones \citep{1974JETP...38..903G}.
The polarization of both modes is close to linear.
The electric field vector of the X-mode photons oscillates perpendicular to the ambient magnetic field direction, while the electric vector of O-mode photons has a component along the field.
The cross sections of the major processes of interaction between radiation and matter are strongly dependent on the polarization mode (see, e.g., \citealt{1984ASPRv...3..197P,2006RPPh...69.2631H}). 
Below the cyclotron resonance, the cross sections for photons of the O-mode tend to be significantly larger in comparison to the cross sections of X-mode photons.
Consequently, one can expect the flux leaving the NS atmosphere to be dominated by the X-mode photons, resulting in a high PD value; however, this value is also dependent on the specific structure of the atmosphere.

As already mentioned in Sect.~\ref{sec:polar}, we did not find any significant difference in the polarization properties of \gro between the two states, with the luminosity differing by a factor of two. 
We note that the geometry of the emitting regions at the NS surface in XRPs is known to depend on accretion luminosity, and in particular is expected to change dramatically with the onset of an accretion column. Indeed, if the luminosity is below the critical level \citep{1976MNRAS.175..395B}, the radiative force in the vicinity of the NS surface is small and the accretion process results in the emitting regions having a hotspot geometry. 
If the luminosity reaches the critical value, the radiative force becomes sufficiently high to stop the accreting material above the NS surface in a radiation-dominated shock, resulting in an extended accretion column above the stellar magnetic poles.
Taking into account the cyclotron energy observed in \gro at around 80 keV and the corresponding surface magnetic field strength $\sim 10^{13}\,{\rm G}$, one would expect the critical luminosity $L_{\rm crit}$ to be a few times $10^{37}\,{\rm erg\,s^{-1}}$ \citep{2015MNRAS.447.1847M}. 
This is an order of magnitude higher than the observed luminosity level in both observations. We therefore expect the radiative force in the vicinity of the NS surface to have no affect on the dynamics of the accretion flow, and expect the X-ray photons to be emitted from hotspots in both \ixpe observations. Therefore, the lack of dramatic changes in the emission region geometry is not surprising.

At subcritical mass-accretion rates, the flow is decelerated in the atmosphere of the NS via Coulomb collisions, resulting in an inverse temperature profile with hotter upper layers and a cooler underlying atmosphere \citep{1969SvA....13..175Z,2018A&A...619A.114S}. 
The typical braking distance of the accretion flow in the atmosphere depends on the flow velocity \citep{1995ApJ...438L..99N}, which is similar in the two states of \gro, and is not affected by the mass-accretion rate. Therefore, the relative distribution of energy release in the atmosphere is expected to be very similar. The absolute value of the local temperature can be different due to different levels of local energy release. On the other hand, considering that the accretion luminosity in the two observations changes by only a factor $\sim2$, the local temperature is also not expected to change significantly (i.e., by a factor of only $2^{1/4}$).  For this reason, we do not expect significant variations in polarization in the two observed states.

Detailed analysis of the radiative transfer problem in the atmosphere of a NS shows that the polarization composition of X-ray flux leaving the atmosphere is strongly affected by the temperature structure and relative contribution of magnetized vacuum and plasma to the dielectric tensor of the medium.
At a certain optical depth in the atmosphere of the NS, the contributions from magnetized vacuum and plasma become comparable, which leads to the mixing of polarization modes, a phenomenon called vacuum resonance.
It appears that the optical depth of the vacuum resonance in the atmosphere influences the final contribution of X- and O- modes to the X-ray energy flux leaving the NS atmosphere. 
In particular, the low PD observed in both states of \gro can be explained if the position of the vacuum resonance and corresponding mode conversion in the atmosphere is located in the transition region where there are strong temperature and mass density gradients, that is, at the border between the overheated upper layer and colder underlying atmosphere (see details in \citealt{Doroshenko22}). 

The correlation between the PD and the flux observed during the pulsation period (see Fig.\,\ref{fig:ixpe-st.pd.pa}) agrees with the expected inverse temperature profile when the accretion luminosity is well below the critical value. 
Under this condition, the beam pattern of the X-ray radiation emitted at the magnetic poles is suppressed along the normal to the stellar surface (because hotter upper layers contribute more to the flux leaving the atmosphere at a larger angle to the normal), while the PD tends to increase with the angle between the normal and the direction along which the photon leaves the atmosphere because of the stronger dependence of cross sections on X-ray polarization at the larger angles \citep{2021MNRAS.503.5193M,2021A&A...651A..12S}. 
When the mass-accretion rate is close to the critical one, as in the case of Cen~X-3, the beam pattern can already be affected by the scattering of X-ray photons by the accretion flow above the NS magnetic poles, leading to the apparent anti-correlation between the PD and the pulsed flux (see Discussion section in \citealt{2022ApJ...941L..14T}).

\begin{figure*}
\centering
\includegraphics[width=0.65\linewidth]{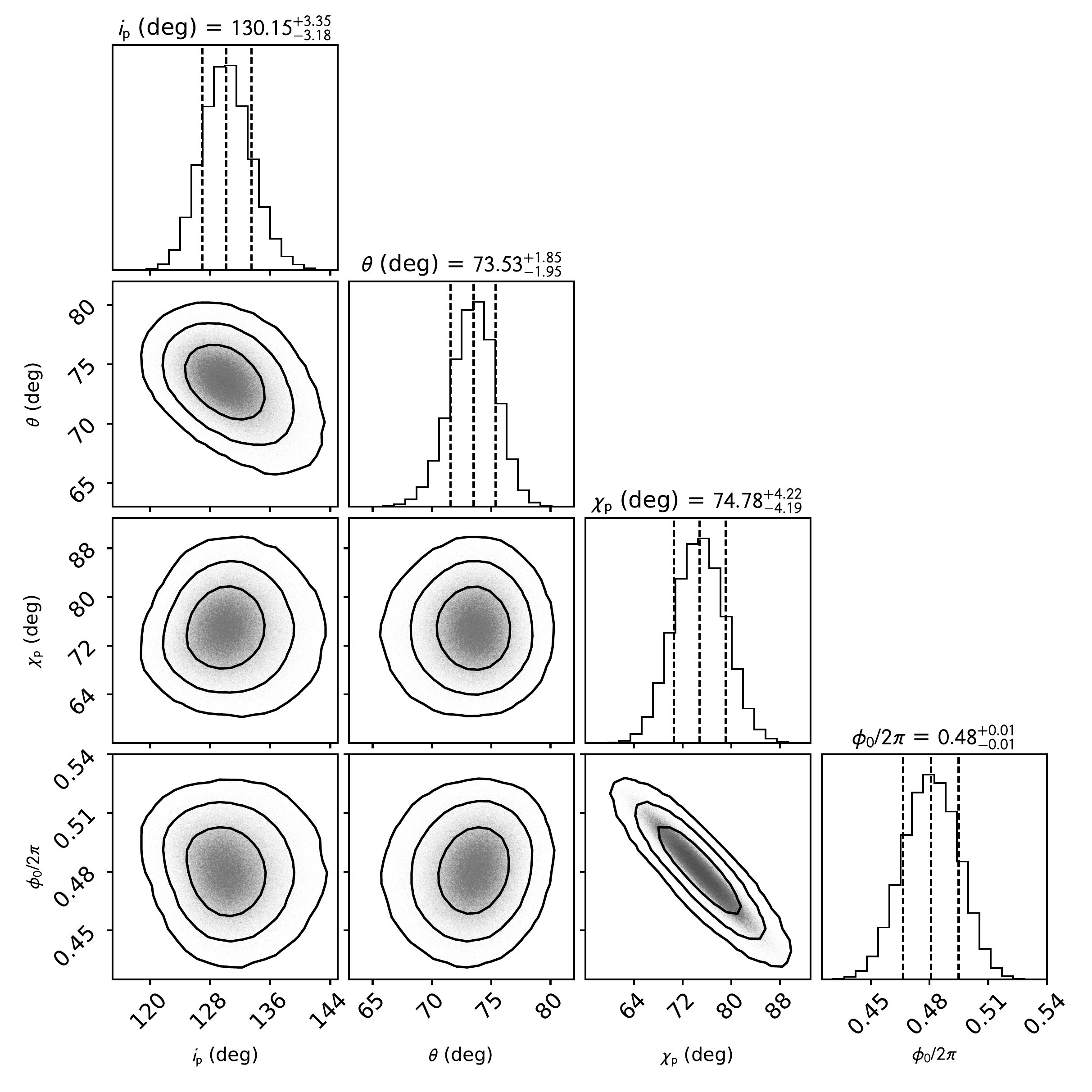}
\caption{Corner plot of the posterior distribution for the RVM parameters for the pulsar geometry obtained using the PA values from the phase-resolved spectro-polarimetric analysis of the combined data. The two-dimensional contours correspond to 68.27\%, 95.45\%, and 99.73\% confidence levels. The histograms show the normalized one-dimensional distribution for each parameter derived from the posterior samples. The mean value and 1$\sigma$ confidence interval for the derived parameters are presented above the corresponding histogram (dashed lines).}
\label{fig:emcee}
\end{figure*}

\subsection{Geometry of the system}
\label{sec:geom}

In order to determine the geometrical parameters of the pulsar, we followed the procedures described in \citet{Doroshenko22} and \citet{2022ApJ...941L..14T}. Namely, we fitted the spin-phase variations of the PA of \gro with the rotating-vector model \citep[RVM;][]{1969ApL.....3..225R,Poutanen20RVM}. 
The applicability of this model ---which assumes the dipole configuration of the NS magnetic field--- to XRPs, where the magnetic field near the star can be more complicated, has been discussed in several papers \citep[e.g.,][]{unbinned,2022Sci...378..646T}. 
Specifically,  the applicability is justified by the birefringence properties of vacuum \citep{1978PAZh....4..214G} causing the radiation propagation in two normal modes until the polarization-limiting radius \citep{1952RSPSA.215..215B,2002PhRvD..66b3002H,2018Galax...6...76H}. 
For a typical XRP, this radius is about 20 stellar radii ($\sim$250~km), which is much larger than the star and therefore we expect the field configuration there to be dipolar. 

If radiation escapes in the O-mode, the PA can be described by the RVM, following the expression  \citep{Poutanen20RVM} 
\begin{equation} \label{eq:pa_rvm}
\tan (\mbox{PA}\!-\!\chi_{\rm p})\!=\! \frac{-\sin \theta\ \sin (\phi-\phi_0)}
{\sin i_{\rm p} \cos \theta\!  - \! \cos i_{\rm p} \sin \theta  \cos (\phi\!-\!\phi_0) } 
,\end{equation} 
where $i_{\rm p}$ is the pulsar inclination (i.e., the angle between the pulsar spin vector and the line-of-sight), $\chi_{\rm p}$ is the position angle of the pulsar spin axis, $\theta$ is the magnetic obliquity (i.e., the angle between the magnetic dipole and the spin axes), $\phi$ is the pulse phase, and $\phi_0$ is the phase when the magnetic pole is closest to the observer.

The pulse-phase dependence of the PA obtained from the spectro-polarimetric analysis of the combined data was fitted with the RVM using the affine invariant Markov chain Monte Carlo ensemble sampler {\sc emcee} package of {\sc python} \citep{2013PASP..125..306F}. 
All the parameters in eq.~(\ref{eq:pa_rvm}) were left free. 
As a result, we obtained accurate estimates of the pulsar inclination, $i_{\rm p}=130\fdg2\pm3\fdg3$, the co-latitude of the magnetic pole, $\theta=73\fdg5\pm1\fdg9$, and the position angle of the pulsar spin, $\chi_{\rm p}=\chi_{\rm p,O}=74\fdg8\pm4\fdg2$ (see Figures~\ref{fig:ixpe-st.pd.pa} and \ref{fig:emcee}). 
However, the pulsar spin-axis position angle is not determined uniquely and can be  directed in the opposite direction at $\chi_{\rm p}=\chi_{\rm p,O}+180\degr=254\fdg8\pm4\fdg2$ because only the orientation of the polarization plane can be measured. 
Also, if radiation escapes in the X-mode, then the pulsar spin is oriented at $\chi_{\rm p}=\chi_{\rm p,X}=\chi_{\rm p,O}\pm90\degr$.
The obtained  geometrical parameters were verified using the RVM fit to the unbinned Stokes parameters on a photon-by-photon basis as outlined in \citet{unbinned} and \citet{2021AJ....162..134M}. 
We find that the PD and PA obtained in this way is nearly identical to those shown in Figure~\ref{fig:ixpe-st.pd.pa}. 

Furthermore, the inclination of the pulsar estimated from \ixpe data ($i_{\rm p}\simeq130\degr$) appears to be close to the orbital inclination of $\sim144\degr$ estimated by \citet{2007MNRAS.378.1427C} based on an analysis of the H$\alpha$ emission line profile from a circumbinary Be disk.\footnote{The actual value of inclination reported by \citet{2007MNRAS.378.1427C} is $\sim36\degr$, but this cannot be distinguished from $180\degr-36\degr=144\degr$.} 
Although the accuracy of the latter estimate might be debated, it indicates that the accretion torques had sufficient time to align the spin and orbital axes. Very rough estimates result in alignment timescales of the order of $\lesssim10^5$ years for a NS with a strong magnetic field \citep[see eq.~16 in][]{2021MNRAS.505.1775B}.

It is interesting to note that the resulting determination of the co-latitude of the magnetic pole points to a very high value of $\theta\approx75\degr$, which tells us that \gro is an almost orthogonal rotator. 
This finding is in stark contrast with results for Cen X-3 and Her X-1, where the magnetic obliquity was much lower, at about 16\degr\ \citep{Doroshenko22, 2022ApJ...941L..14T}. 
Observations of other XRPs with \ixpe, that is, Vela X-1  \citep{Forsblom2023} and X Persei \citep{Mushtukov23}, also imply either low or high values of magnetic obliquity. 
These results are in fact in line with some studies predicting a bimodal distribution of the magnetic obliquity $\theta$ peaking around 0 and 90 deg in the case of isolated NSs \citep[e.g.,][]{2018MNRAS.481.4169L}. 
However, we note that in the case of accreting NSs, the predictions are less specific depending on the accretion mechanism \citep{2021MNRAS.505.1775B}. 
More observations of XRPs with \ixpe will allow us to verify models of the axis alignment of accreting pulsars.

\section{Summary}
\label{sec:sum}

The results of our study can be summarized as follows:

\begin{enumerate}
\item \gro was observed by \ixpe during the Type I outburst in November 2022 in two states, which differ in flux level by a factor of 2.

\item Both the energy-binned and spectro-polarimetric  analyses of the phase-averaged data reveal a significant average polarization of the source with a PD of $\sim$3.9\% regardless of the source flux.

\item The pulse-phase resolved analysis reveals a correlation between the flux and the PD, and suggests that a strong variation of the PA is responsible for the low average polarization from the source. The results obtained in the two luminosity states are consistent within the uncertainties. We also find an indication that the PD is  positively correlated with energy.

\item The observed variations of the PA are well described in the framework of the rotating-vector model. The corresponding inclination of the pulsar is about 130\degr, and the position angle of the pulsar spin is  $\sim$75\degr\ (or 255\degr) if radiation escapes from the surface in the O-mode, or $\sim$165\degr\ (or $-15\degr$) if the X-mode dominates. In all cases, the magnetic obliquity is found to be very large $\sim$75\degr, implying that \gro is  a nearly orthogonal rotator. 

\item The observed pulsar inclination appears to agree with the estimated orbital inclination of the system, suggesting that the pulsar spin is close to being aligned with the orbital axis. This indicates that, in spite of a strong natal kick received by the NS, the accretion torques (even though acting sporadically) had sufficient time to align the spin and orbital angular momenta.

\item The relatively low polarization detected from \gro as well as the finding that the polarization properties are  independent of the mass accretion rate can be explained in the framework of the model of an overheated NS atmosphere. 
\end{enumerate}

Detecting changes in the polarization properties with mass accretion rate requires observations of an XRP during giant Type II outburst. Such observations were performed by \ixpe in February 2023 for the Be/XRP LS V +44 17/RX J0440.9+4431, where strong variations of the PD and PA in different luminosity states were discovered (Doroshenko V. et al., in prep.). 

\begin{acknowledgements}
The Imaging X-ray Polarimetry Explorer (IXPE) is a joint US and Italian mission.  
The US contribution is supported by the National Aeronautics and Space Administration (NASA) and led and managed by its Marshall Space Flight Center (MSFC), with industry partner Ball Aerospace (contract NNM15AA18C).  
The Italian contribution is supported by the Italian Space Agency (Agenzia Spaziale Italiana, ASI) through contract ASI-OHBI-2017-12-I.0, agreements ASI-INAF-2017-12-H0 and ASI-INFN-2017.13-H0, and its Space Science Data Center (SSDC) with agreements ASI-INAF-2022-14-HH.0 and ASI-INFN 2021-43-HH.0, and by the Istituto Nazionale di Astrofisica (INAF) and the Istituto Nazionale di Fisica Nucleare (INFN) in Italy.
This research used data products provided by the IXPE Team (MSFC, SSDC, INAF, and INFN) and distributed with additional software tools by the High-Energy Astrophysics Science Archive Research Center (HEASARC), at NASA Goddard Space Flight Center (GSFC). 

We acknowledge support from the Academy of Finland grants 333112, 349144, 349373, and 349906 (SST, JP), the German Academic Exchange Service (DAAD) travel grant 57525212 (VD, VFS), the V\"ais\"al\"a Foundation (SST), the German Research Foundation (DFG) grant \mbox{WE 1312/53-1} (VFS), UKRI Stephen Hawking fellowship and the Netherlands Organization for Scientific Research Veni fellowship (AAM), the Czech Science Foundation project 21-06825X (JS).
\end{acknowledgements}


\end{document}